\documentclass[useAMS,usenatbib]{mn2e}
\usepackage{epsf}
\usepackage{amssymb}
\usepackage[usenames]{color}
\usepackage{subfloat}

\def\plotone#1{\centering \leavevmode
\epsfxsize=\columnwidth \epsfbox{#1}}

\def\plottwo#1#2{\centering \leavevmode
\epsfxsize=1.03\columnwidth \epsfbox{#1} \hfil
\epsfxsize=1.03\columnwidth \epsfbox{#2}}

\def\plottwos#1#2{\centering \leavevmode
\epsfxsize=.8\columnwidth \epsfbox{#1} \hfil
\epsfxsize=.8\columnwidth \epsfbox{#2}}

\def\kms{\,{\rm km~s}^{-1}}

\newcommand{\be}{\begin{equation}}
\newcommand{\ee}{\end{equation}}

\def\disp {\displaystyle}
\title[ Comparison of simple mass estimators]{Comparison of simple mass estimators for slowly rotating elliptical galaxies}
\author[Lyskova et al.]{N.~Lyskova$^{1,2}$, J.~Thomas$^{3,4}$, E.~Churazov$^{1,2}$,
S.~Tremaine$^{5}$, T.~Naab$^{1}$
\newauthor \\
$^1$ Max-Planck-Institut f\"ur Astrophysik, Karl-Schwarzschild-Strasse 1, 85741
Garching, Germany\\
$^2$ Space Research Institute (IKI), Profsoyuznaya 84/32, Moscow 117810, 
Russia\\
$^3$ Max-Planck-Institut f\"{u}r Extraterrestrische Physik, P.O.\ Box 1312, 85741 Garching, Germany\\
$^4$ Universit\"ats-Sternwarte M\"unchen, Scheinerstr. 1, D-81679 M\"unchen, Germany \\
$^5$ Institute for Advanced Study, Princeton, NJ 08540, USA
}

\begin{document}

\pagerange{\pageref{firstpage}--\pageref{lastpage}}
\pubyear{2014}

\maketitle

\label{firstpage}
\begin{abstract}

We compare the performance of mass estimators for elliptical galaxies that rely on the directly observable surface brightness and velocity dispersion profiles, without invoking computationally expensive detailed modeling. These methods recover the mass at a specific radius where the mass estimate is expected to be least sensitive to the anisotropy of stellar orbits.  
One method (\citealt{Wolf.et.al.2010}) uses the total luminosity-weighted velocity dispersion and evaluates the mass at a 3D half-light radius $r_{1/2}$, i.e., it depends on the {\it global} galaxy properties. Another approach (\citealt{Churazov.et.al.2010}) estimates the mass from the velocity dispersion at a radius $R_2$ where the surface brightness declines as $R^{-2}$, i.e., it depends on the {\it local}  properties.
We evaluate the accuracy of the two methods for analytical models, simulated galaxies and real elliptical galaxies that have already been modeled by the Schwarzschild's orbit-superposition technique. Both estimators recover an almost unbiased circular speed estimate  with a modest RMS scatter ($\lesssim 10 \%$). Tests on analytical models and simulated galaxies indicate that the local estimator has a smaller RMS scatter than the global one. We show by examination of simulated galaxies that the projected velocity dispersion at $R_2$ could serve as a good proxy for the virial galaxy mass. For simulated galaxies the total halo mass scales with $\sigma_p(R_2)$ as $\disp M_{vir} \left[M_{\odot}h^{-1}\right] \approx 6\cdot 10^{12} \left( \frac{\sigma_p(R_2)}{200\, \rm km\, s^{-1}} \right)^{4}$ with RMS scatter $\approx 40 \%$.

\end{abstract}


%

\sloppypar

\section{Introduction}

Galaxy masses play a key role in our understanding of their formation and evolution. While observations of disc rotation curves allow the determination of spiral galaxy masses directly, the situation with early-type galaxies is more complex due to the lack of mass tracers on known orbits. Approaches of different levels of sophistication and generality have been developed for mass determination of elliptical galaxies. Dynamical modeling using the orbit-superposition method is considered to be the state-of-the-art technique and allows one to recover both the galaxy gravitational potential and the orbital structure with an accuracy of $\lesssim 15\%$ \citep{Thomas.et.al.2005, Krajnovic.et.al.2005}. Schwarzschild modeling is now widely used for mass measurements of supermassive black holes, for determination of the total mass profile and its decomposition into luminous and dark matter components as well as for constraining the orbital structure \citep[e.g.,][]{Gebhardt.et.al.2003, 
Cappellari.et.al.2006, Thomas.et.al.2007b, Thomas.et.al.2009, Thomas.et.al.2011, McConnell.et.al.2012, McConnell.et.al.2013, Rusli.et.al.2013}.  As such an approach  only makes sense with  
high quality observational data allowing the determination of the high order line-of-sight velocity moments (namely, in addition to velocity and projected velocity dispersion also the third and the fourth order Gauss-Hermite moments \citep[e.g.,][]{Gerhard.1993,van.der.Marel.Franx.1993} or better yet the complete line-of-sight velocity distribution), it can only be applied to nearby galaxies. Moreover, numerical experiments show that due to intrinsic degeneracies not all the model parameters can be uniquely constrained even from the best available integral-field stellar kinematics \citep[e.g.,][]{Thomas.et.al.2007a,van.den.Bosch.van.de.Ven.2009}.

Large observational surveys at lower resolution have become a major tool for galaxy studies as they facilitate a number of  statistical investigations of galaxy properties. Here, exact galaxy mass determinations at different redshifts are critical for galaxy formation studies and for tracing the assembly of galaxy mass over time. With low-resolution information about galaxy photometry and kinematics, the usage of detailed dynamical modeling is impractical/not justified. It is desirable to have simple and robust techniques based on the most easily accessible observables that provide mass estimates with a modest and known scatter.

Recently two simple mass estimation methods have been suggested (\citealt{Churazov.et.al.2010} and \citealt{Wolf.et.al.2010}) which evaluate masses at a special radius from the surface brightness and projected velocity dispersion profiles without detailed modeling. Although these approaches recover the mass at some particular radius only and not the mass distribution within that radius, these estimates could be used
(i)   to determine the non-thermal contribution to the total gas pressure when compared with the X-ray mass estimate at the same radius;
(ii)  to derive a slope of the mass profile when combined with the mass estimate from strong lensing;
(iii) as a virial mass proxy.

Here we compare the performance of simple mass estimators on analytical models, simulated galaxies, and also on a set of real elliptical galaxies with high-quality kinematical data and Schwarzschild modeling results. The paper is organized as follows.
In Section~\ref{sec:mass_app}, we provide a brief description of the simple mass estimators. We present results of the tests  on analytical models and a sample of simulated galaxies in Section~\ref{sec:tests} and on real elliptical galaxies  in Section~\ref{sec:art}. The possibility of using these estimates as a proxy for the virial mass is discussed in  Section~\ref{sec:proxy}. A summary of the accuracy and biases of the methods and conclusions are given in Section~\ref{sec:conclusion}.

\section{Mass approximation formulae}
\label{sec:mass_app}

High resolution observations of the stellar surface brightness and line-of-sight velocity dispersion profiles of galaxies are the basis of various dynamical modeling techniques of different levels of sophistication and generality.
Unfortunately, all dynamical methods probably suffer from a common problem: even with the best available kinematics and photometry they are usually unable to uniquely resolve several degeneracies \citep[e.g.,][]{Dejonghe.Merritt.1992}. The key problems are (i) the degeneracy  between the total mass profile and velocity anisotropy $\beta$ \citep{Binney.Tremaine.2008} and (ii) the deprojection of the surface brightness profile $I(R)$ into a three-dimensional luminosity density for axisymmetric or triaxial systems  \citep[e.g.,][]{Rybicki.1987, Gerhard.Binney.1996}. 

The (scalar) virial theorem provides a straightforward way to estimate the mass (or circular velocity) of an elliptical galaxy. For an isolated stationary spherical system in a logarithmic gravitational potential $\Phi(r)=V_c^2\ln(r) + \rm const$ the circular velocity $V_c$ is related to the average (luminosity-weighted) line-of-sight velocity dispersion as 

\begin{eqnarray}
V_c^2=3\langle \sigma_p^2 \rangle=3\,\frac{\int_0^\infty \sigma_p^2(R)I(R)R\,dR}{\int_0^\infty I(R) R\,dR }.
\label{eq:virial}
\end{eqnarray}

The mass is then equal to $ M(<r) = rV_c^2/G$, where $G$ is the gravitational constant. Under the above ideal assumptions,  this approach is rigorously independent of the velocity anisotropy. However, in practice, observed galaxies are not guaranteed (i) to have spectroscopic data over the full extent of the systems, (ii) to be spherically symmetric and/or (iii) to have a stellar and dark matter distribution that sum up to an exactly logarithmic gravitational potential. 

To weaken some of these assumptions it is common to use the spherical Jeans equation which relates the velocity anisotropy parameter $\beta(r)$, the  luminosity density of the stars $j(r)$ and their radial velocity dispersion $\sigma_r(r)$: 

\begin{equation}
\displaystyle {d\over dr} \left(j\sigma_r^2 \right)+2\frac{\beta}{r}j\sigma_r^2=-j{d\Phi\over dr},
\label{eq:Jeans}
\end{equation} 
where $\beta(r)~=1~-~\sigma_{\theta}^2/\sigma_{r}^2$ ($\sigma_{\theta}(r)$ is the tangential velocity dispersion; and in spherical symmetry  $\sigma_{\theta} \equiv \sigma_{\phi}$).

In principle, one can derive $M(<r)$ from the Jeans equation linking the 3D quantities $j(r)$ and $\sigma_r(r)$ to the observable surface brightness $I(R)$ and projected velocity dispersion $\sigma_p(R)$ via

\begin{equation}
\displaystyle I(R)=2\int_R^\infty\!\! {jr\,dr\over\sqrt{r^2-R^2}},
\label{eq:Abel}
\end{equation}

\begin{equation}
\sigma_p^2(R) I(R)=2\int_R^\infty\!\! \left(1-\frac{R^2}{r^2}\beta\right){j\sigma_r^2r\,dr\over\sqrt{r^2-R^2}}
\label{eq:Abel2}
\end{equation} 
for any given anisotropy profile.
Unfortunately observational data alone do not allow one to constrain $\beta(r)$ without invoking sophisticated detailed modeling and additional observational constraints. From considering relation (\ref{eq:Abel2}) one might expect to find some characteristic radius $R_{char}$ where the uncertainty in the circular speed estimate arising from the unknown anisotropy is minimal, thus mitigating the mass-anisotropy degeneracy at the expense of losing spatial resolution/information. The existence of such a radius was first noted in \cite{Richstone.Tremaine1984}. In this case the mass estimator can be expressed in the following form:

\begin{equation}
\displaystyle V_c^2(R_{char})=k\sigma_p^2,
\label{eq:genform}
\end{equation} 
where $\sigma_p$ is some measure of the velocity dispersion designed so that  the dependence of $k$ on the orbital structure is as small as possible.

This idea is the basis of the simple dynamical mass scaling relations discussed in detail by \cite{Churazov.et.al.2010} and \cite{Wolf.et.al.2010}. In the remainder of the paper we concentrate on these two mass estimators and test their performance.

\subsection{Local estimator}
\label{subsec:EC}

The Churazov et al. estimator is derived from the stationary non-streaming spherical Jeans equation under the assumption of a logarithmic gravitational potential $\Phi(r)=V_c^2\ln(r)+\rm const$. In this case projected velocity dispersion profiles can be analytically derived for  isotropic ($\beta=0$), radial ($\beta=1$) and circular ($\beta\rightarrow -\infty$) stellar orbits:

\[
\sigma^{\rm iso}_p(R) = V_c(r) \frac{1}{\sqrt{1+\alpha+\gamma}} 
\]
\be
\sigma^{\rm circ}_p(R) = V_c(r) \sqrt{\frac{\alpha}{2 (1+\alpha+\gamma)}}
\label{eq:main} 
\ee
\[
\sigma^{\rm rad}_p(R) = V_c(r)\frac{1}{\sqrt{\left(\alpha+\gamma\right
  )^2+\delta-1}}, 
\]
where 
\be
\alpha\equiv-\frac{d\ln I}{d\ln R}, \ \ \gamma\equiv -\frac{d\ln
  \sigma_p^2}{d\ln R},\ \ \delta\equiv \frac{d^2\ln[I\sigma_p^2]}{d
  (\ln R)^2}.
\label{eq:agd}
\ee 

These equations are only exact for a logarithmic potential, that is, if the circular speed is independent of radius. 
However, on scales where the circular speed varies sufficiently slowly the projected velocity dispersions for $\beta = 0, -\infty, 1$ approximately follow relations (\ref{eq:main}), which allows one to invert eq.~(\ref{eq:main}) and form estimates of $V_c$ for given anisotropy values ($V^{\rm iso}_c$,  $V^{\rm circ}_c$ and  $V^{\rm rad}_c$) via local properties of $I(R)$ and $\sigma_p(R)$.   

For typical galaxy potentials, there exists a 
characteristic radius $R_{char}$ (called $R_{\rm sweet}$ by \citealt{Churazov.et.al.2010}), where the circular speed 
estimate does not strongly depend on  $\beta$. Ideally, this 
radius would be defined as the point of intersection
of $V^{\rm iso}_c$,  $V^{\rm circ}_c$ and  $V^{\rm rad}_c$, i.e., where circular speed estimates for different $\beta$ give exactly the same values. In real galaxies $V^{\rm iso}_c$,  $V^{\rm circ}_c$ and  $V^{\rm rad}_c$ are not guaranteed to intersect at one radius and $R_{\rm sweet}$ is defined as the radius where these three curves are most close to each other, i.e., where the circular speed estimate is largely independent of $\beta$. In this paper we calculate $R_{\rm sweet}$ as the radius at which the standard deviation among the three curves is minimized.
As seen from the equations above, $R_{\rm sweet}$ depends only on the {\it local} properties of the observed surface brightness $I(R)$ and projected velocity dispersion $\sigma_p(R)$ profiles. 

For massive elliptical galaxies the spatial variation of $\sigma_p(R)$ is typically much smaller than that of $I(R)$, i.e., $\gamma \ll \alpha$ and $\delta \ll \alpha$. Hence, the estimates of circular speed derived from (inverted) eq.~(\ref{eq:main}) can be approximated by:

\[
V^{\rm iso, s}_c=\sigma_{p}(R)  \sqrt{\alpha+1} 
\]
\be
V^{\rm circ, s}_c=\sigma_{p}(R)   \sqrt{2\frac{\alpha+1}{\alpha}}
\label{eq:agd_simple} 
\ee
\[
V^{\rm rad, s}_c=\sigma_{p}(R)  \sqrt{\alpha^2-1}. \\
\]

Note that for $\alpha = 2$  the relation between the circular speed and the observed velocity dispersion is the same for isotropic, circular, and radial orbits and in fact this equality extends to all systems with constant anisotropy $\beta$  \citep{Gerhard.1993}, i.e., for the ideal case of a nearly flat projected velocity dispersion profile, the characteristic radius $R_{sweet}=R_{2}$ where the surface brightness declines as $R^{-2}$. For general spherical models $R_{\rm sweet}$ is expected to lie not far from $R_{2}$ which, in turn, is close to the projected half-light radius (sometimes called the effective radius) $R_{1/2}$. Thus, in practice, one can use $R_2$ instead of $R_{\rm sweet}$ as the characteristic radius for the circular speed estimate. In this case the sweet radius is derived from the logarithmic slope of the surface brightness profile only. So we may consider the following three local circular speed estimators
\begin{itemize}
\item {\bf estimator L1:} $V_c^{\rm iso}(R_{\rm sweet})$ where $V_c^{\rm iso}$ is given by the first of eq.~(\ref{eq:main}) and $R_{\rm sweet}$ is the radius at which the three estimators of $V_c$ in eq.~(\ref{eq:main}) are closest to one another.

\item {\bf estimator L2:} $V_c^{\rm iso}(R_2)$ where $V_c^{\rm iso}$ is the same as in L1 and $R_2$ is the radius at which the logarithmic slope of the surface brightness distribution is $-2$ ($\alpha=2$ in eq.~(\ref{eq:agd})).

\item {\bf estimator L3:} $V_c^{\rm iso,s}(R_2)$ where $V_c^{\rm iso,s}$ is given by the first of eq.~(\ref{eq:agd_simple}) and $R_2$ is the same as in L2.

\end{itemize}

\subsection{Global estimator}
\label{subsec:W}

The mass estimator suggested by \cite{Wolf.et.al.2010} is derived from the spherical Jeans equation combined with the scalar virial theorem under the assumption of a fairly flat line-of-sight velocity dispersion profile $\sigma_p(R)$.
Wolf et al. show that the mass uncertainty arising from the unknown anisotropy of stellar orbits $\beta$ is minimized at the radius $r_{3}$ where the 3D luminosity density profile $j(r)$ decays as $r^{-3}$.
Within this radius the galaxy mass can be inferred from the luminosity-weighted projected velocity dispersion profile $\langle \sigma_p^2 \rangle$, averaged from $0$ to $\infty$:

\be
M(<r_3) \simeq 3G^{-1}\langle \sigma_p^2 \rangle r_3,
\label{eq:w1}
\ee
or in terms of the circular speed (hereafter, the global {\bf estimator G1})

\be
V_c^2 \left( r_3 \right) \simeq 3\langle \sigma_p^2 \rangle.
\label{eq:w1.1}
\ee
As seen from these equations, the mass estimator depends on the {\it global} properties of the observed surface brightness $I(R)$ and the projected velocity dispersion $\sigma_p(R)$ profiles.

As discussed in \cite{Wolf.et.al.2010}, for a wide range of stellar distributions (e.g. for exponential, Gaussian, King, S\'{e}rsic profiles) $r_3$ is close to the 3D half-light radius $r_{1/2}$, which is in turn approximately equal to $\frac{4}{3} R_{1/2}$, where $R_{1/2}$ is the projected half-light radius. So the circular speed estimate can also be expressed as 

\be
V_c^2 \left( r_{1/2} \right) \simeq 3\langle \sigma_p^2 \rangle
\label{eq:w3}
\ee
{\bf (estimator G2)} or

\be
V_c^2 \left( \frac{4}{3} R_{1/2} \right) \simeq 3\langle \sigma_p^2 \rangle
\label{eq:w3}
\ee
{\bf (estimator G3)}.

These mass estimators are intended for use with galaxies that (i) are spherical, (ii) are non-rotating, (iii) have spatially resolved kinematics over the entire galaxy and (iv) have a projected velocity dispersion that does not vary with radius.

\begin{table*}
\centering
\caption{\label{tab:description} Main properties of the Churazov et al. (local) and the Wolf et al. (global) estimators. }
\begin{tabular}{lcclccc}
\hline
\rule{0cm}{0.5cm}
Estimator   & Assumptions & Data & Formulae $V_c^2(R_{char}) = k \sigma_p^2$\\[0.1cm]
\hline
\rule{0cm}{0.3cm}
Local       &dynamical equilibrium&log-slope of $I(R)$, &L1: $V_c^2(R_{\rm sweet}) = \left[ 1+\alpha(R_{\rm sweet})+\gamma(R_{\rm sweet}) \right] \cdot \sigma_p^2(R_{\rm sweet})$, \\[0.1cm]
(L1, L2, L3)&spherical symmetry   & $\sigma_p(R_{sweet})$ or $\sigma_p(R_{2})$,   &\hspace{0.5cm} where $R_{sweet}$: $V^{\rm iso}_c \approx V^{\rm circ}_c \approx V^{\rm rad}_c$ (eq.~(\ref{eq:main})).\\[0.1cm] 
            &no rotation          & log-slope of $\sigma_p(R)$, &  L2: $V_c^2(R_{2}) = \left[ 3+\gamma(R_{2}) \right]\cdot \sigma_p^2(R_{2})$,  \\[0.1cm]
            &logarithmic potential & $d^2\ln[I\sigma_p^2]/d(\ln R)^2$& \hspace{0.5cm} where $R_2$: $I(R) \propto R^{-2}$.    \\[0.1cm]
            &                     &                              & L3: $V_c^2(R_2) =3\cdot \sigma_p^2(R_{2})$.  \\[0.1cm]
\hline
\rule{0cm}{0.3cm}
Global      &dynamical equilibrium& $\sigma_p(R)$ over entire galaxy,& G1: $V_c^2(r_3) = 3\langle \sigma_p^2 \rangle$,   \\[0.1cm]
(G1, G2, G3)&spherical symmetry   & deprojection of $I(R)$          & \hspace{0.5cm} where $r_3$: $j(r) \propto r^{-3}$.  \\[0.1cm]
            &no rotation          & or determination of $R_{\rm 1/2}$& G2:  $V_c^2(r_{1/2})= 3\langle \sigma_p^2 \rangle$,       \\[0.1cm]
            &flat $\sigma_p(R)$   &                                 & \hspace{0.5cm} where $r_{1/2}$ is the 3D half-light radius. \\[0.1cm]
            &                     &                                 & G3: $V_c^2\left(\frac{4}{3} R_{1/2}\right)= 3\langle \sigma_p^2, \rangle$ \\[0.1cm]                  
            &                     &                                 & \hspace{0.5cm} where $R_{1/2}$ is the projected half-light radius.      \\[0.1cm]   
\hline
\end{tabular}
\end{table*}

Both the Churazov et al. and the Wolf et al. mass estimators are based on the spherical Jeans equation and only allow the determination of the total mass within a sphere of some particular characteristic radius $R_{char}$, at which the mass estimate is not sensitive to the anisotropy. Despite these similarities, the final equations and expressions for $R_{char}$ look quite different. The crucial difference between these mass estimators is the following: the Churazov et al. $V_c$-estimate depends only on local properties of the observed profiles ($\sigma_p(R)$ and the log-slopes of $I(R)$ and $\sigma_p(R)$), while the Wolf et al. formula requires averaging of the velocity dispersion over the entire extent of the galaxy and the determination of the projected half-light radius, i.e., it depends on the global galaxy properties.
The main properties of the Churazov et al. and the Wolf et al. estimators are summarized in Table~\ref{tab:description}.

\section{Performance tests}
\label{sec:tests}

In order to compare the performance of the local and global approaches, we test them on analytical models, simulated galaxies and real elliptical galaxies, the latter having also been analyzed using Schwarzschild modeling.

\subsection{Analytical models}
\label{subsec:an_mod}

\begin{figure*}
\plottwo{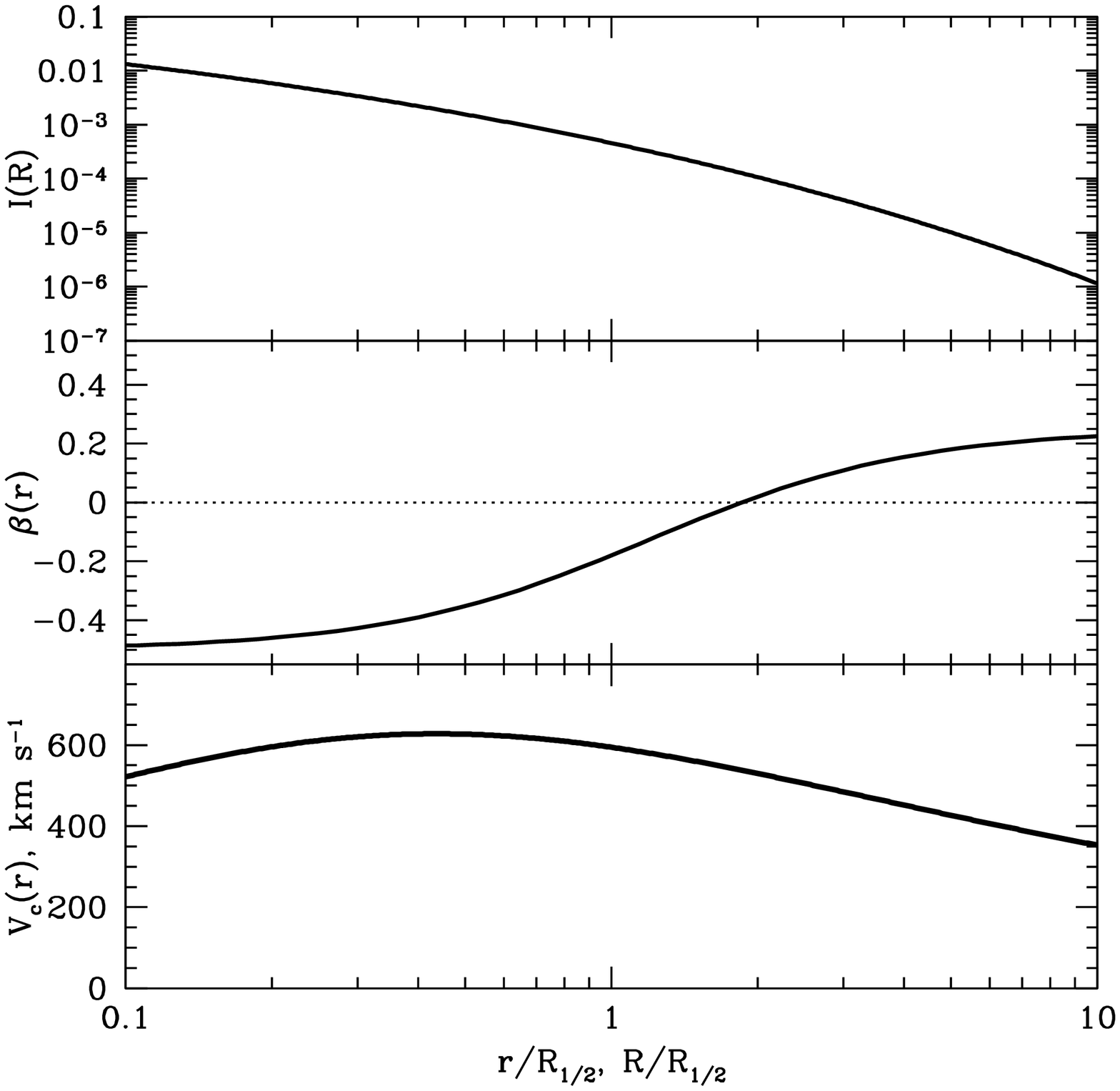}{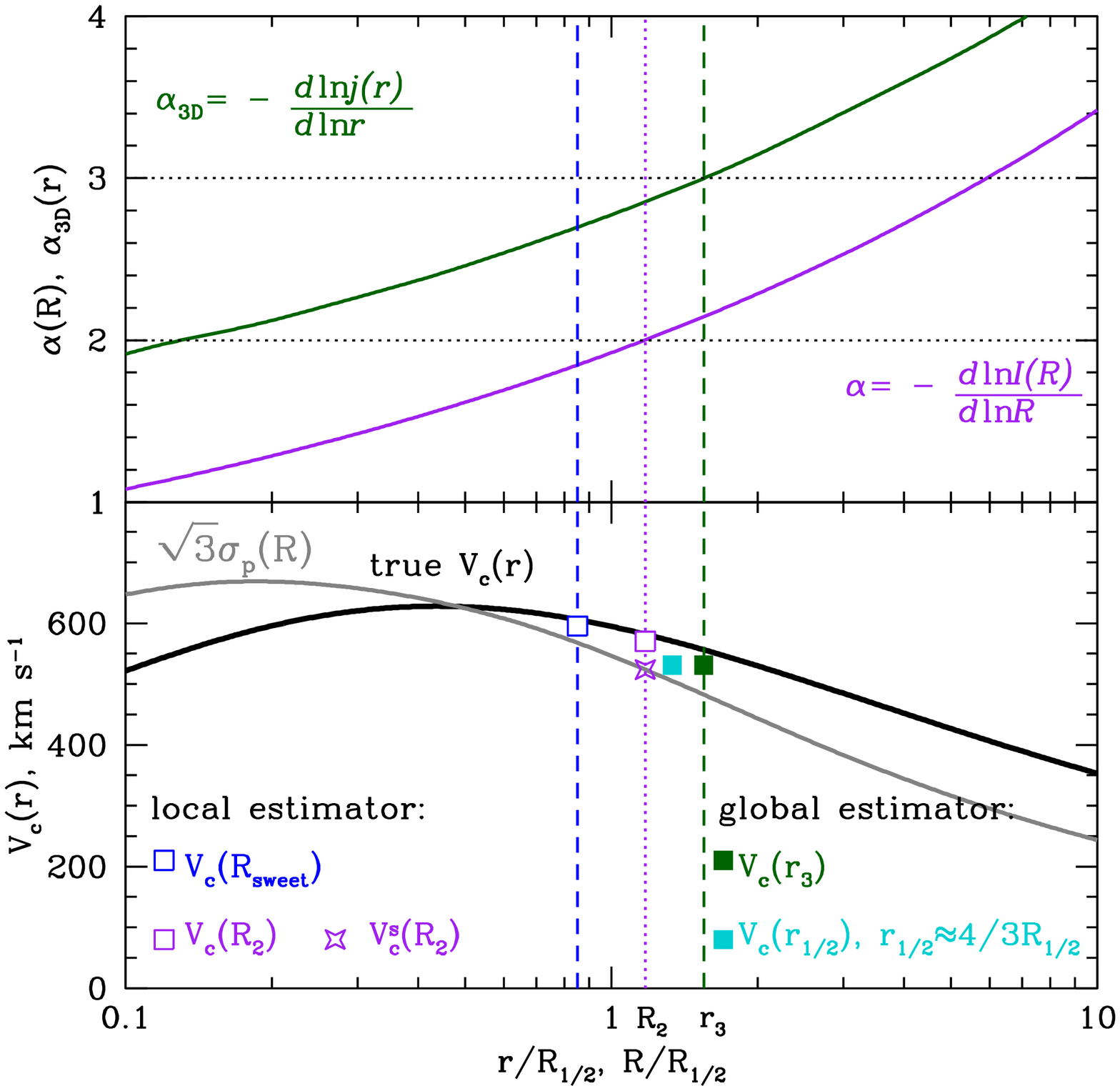}
\caption{Profiles for a typical analytical model (here, $n=4$ S\'ersic model). Left: assumed surface brightness $I(R)$, anisotropy $\beta(r)$ and circular speed $V_c(r)$ as functions of $R/R_{\rm 1/2}$ (or $r/R_{\rm 1/2}$), where $R_{1/2}$ is the projected half-light radius. Right: profiles used to get the simple $V_c$-estimates. Logarithmic slopes of $j(r)$ (obtained from  the spherical deprojection of the surface brightness) and $I(R)$ are shown in the upper panel as dark green and purple curves respectively. $r_3$, $R_{\rm sweet}$ and $R_2$ are marked as dark green dashed, blue dashed and purple dotted lines. The velocity dispersion profile $\sigma_p(R)$ from the spherical Jeans equation is shown as a grey line in the lower panel. The simple circular speed estimates are shown as open and filled symbols of different colors. The filled dark green square is for the global $V_c$-estimate at $r_3$ (G1). In this model $r_{1/2}$ is within $\sim 1\%$ of $\frac{4}{3}R_{1/2}$ so we plot G2 and G3 together as a filled cyan square. The open blue square shows the local estimate at $R_{\rm sweet}$ (L1) and the open purple square at $ R_{2}$ (L2), the open purple star shows the simplified version of the local estimator (L3).   
\label{fig:demo}
}
\end{figure*}

To set up the analytical models we solve the Jeans equation numerically for a set of assumed 3D luminosity density, anisotropy and circular speed profiles. The 3D luminosity density comes from the spherical deprojection of the S\'{e}rsic profile $ I(R) =I(R_{1/2}) \exp\left[-b_{n}\left( (R/R_{1/2})^{1/n}-1\right)\right] $ \citep{Sersic.1968}, where $b_{n}$ obeys  $ \Gamma(2n)=2\gamma(2n,b_n)$. $ \Gamma(2n)$ is the complete gamma function and $\gamma(2n,b_n)$ is the lower incomplete gamma function. We assume an anisotropy $ \beta(r)= \left( \beta_2 r^{c}+\beta_{1}r_a^{c} \right)/ \left( r^{c}+r_a^{c} \right)$. Here $c$ is a concentration parameter and $r_a$ is some characteristic anisotropy radius,  where the anisotropy changes from its central value, $\beta_1$, to its outer value, $\beta_2$. In real early-type galaxies the distribution of stellar orbits is believed to be mildly tangential at the very center \citep{Thomas.et.al.2014}, close to isotropic at small and moderate distances from the center and more radial in the outer parts \citep[e.g.,][]{Gerhard.et.al.2001, Thomas.et.al.2009}.
Thus we vary $\beta_1$ from $-0.7$ to $0.0$ and $\beta_2$  from $0.1$ to $0.5$. We vary $r_a$ and $c$ in the range  $0.1R_{1/2} \le r_a \le 3.2R_{1/2}$ and $0.1 \le c \le 5.1$, respectively.

The circular speed is parametrized as 
\be
V^2_c(r) \propto V^2_{1}\frac{r^2}{r^2+r_c^2} + V^2_{2}\frac{\ln(1+r/r_s)-r/(r+r_s)}{r/r_s}
\label{eq:vc}
\ee
to mimic measured  $V_c$-profiles \citep[e.g.,][]{Gerhard.et.al.2001, Breddels.et.al.2013}. In other words, the gravitational potential of a model spherical galaxy is assumed to be a combination of the cored logarithmic potential \citep{Binney.Tremaine.2008} and the potential
which is similar in shape to a Navarro-Frenk-White potential \citep{NFW}.  Let us note that this is a simple analytical representation of the circular velocity and the free parameters are chosen to cover the wide range of  observed shapes and to probe different parts  of real circular speed profiles (rising, roughly flat and decreasing $V_c$). The parameter $r_s$ varies from $0$ to $90$ half-light radii, $r_c$ changes from $0$ to $12R_{1/2}$.

The luminosity-weighted average of the projected velocity dispersion profile $ \langle \sigma_p^2 \rangle $ is calculated over $[0.1R_{1/2};10R_{1/2}]$.
Figure~\ref{fig:demo} illustrates all steps of the analysis and shows typical profiles considered for analytical models.

\subsubsection{Ideal models}
\label{subsubsec:ideal}

\begin{figure*}
\plottwo{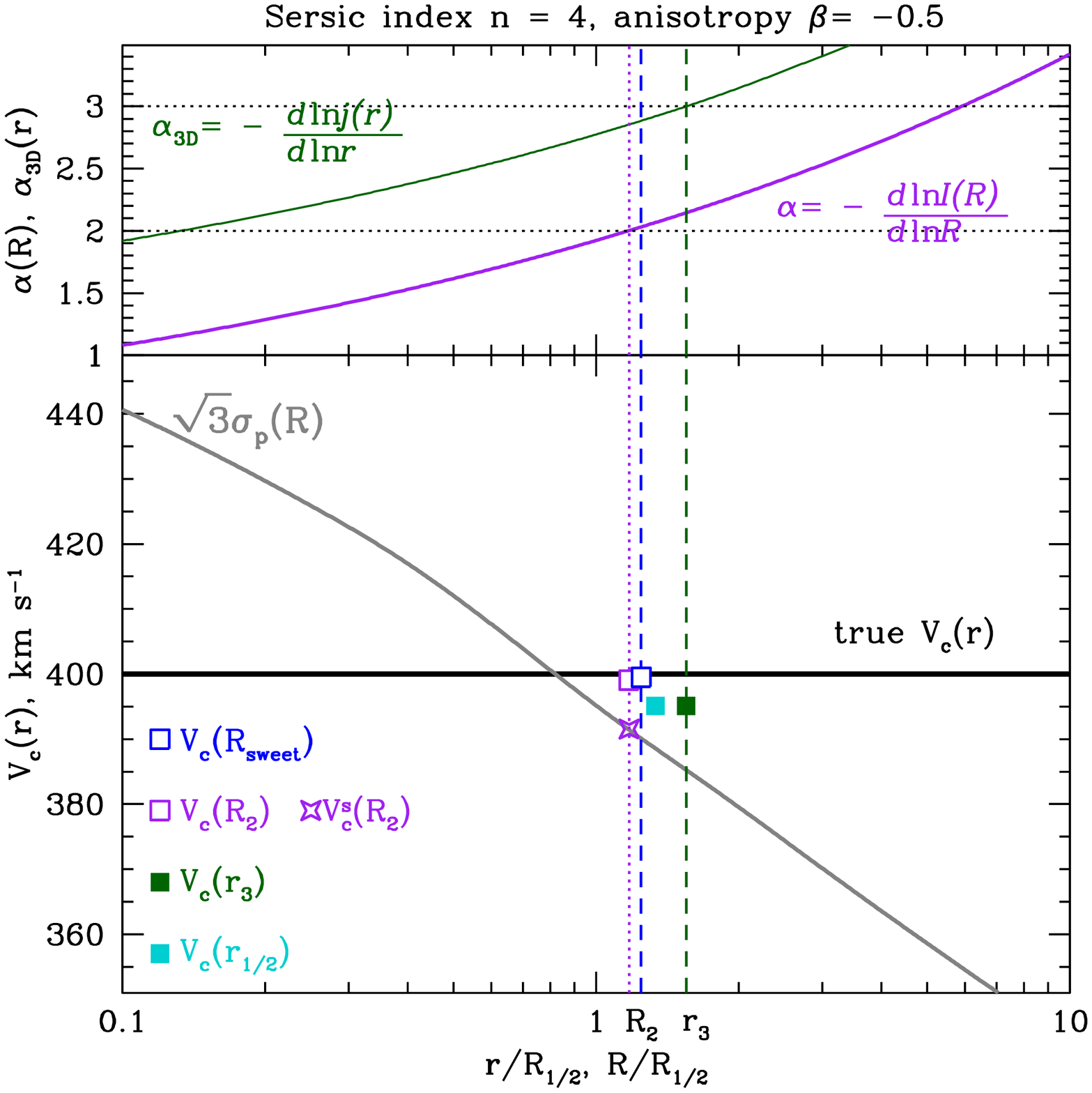}{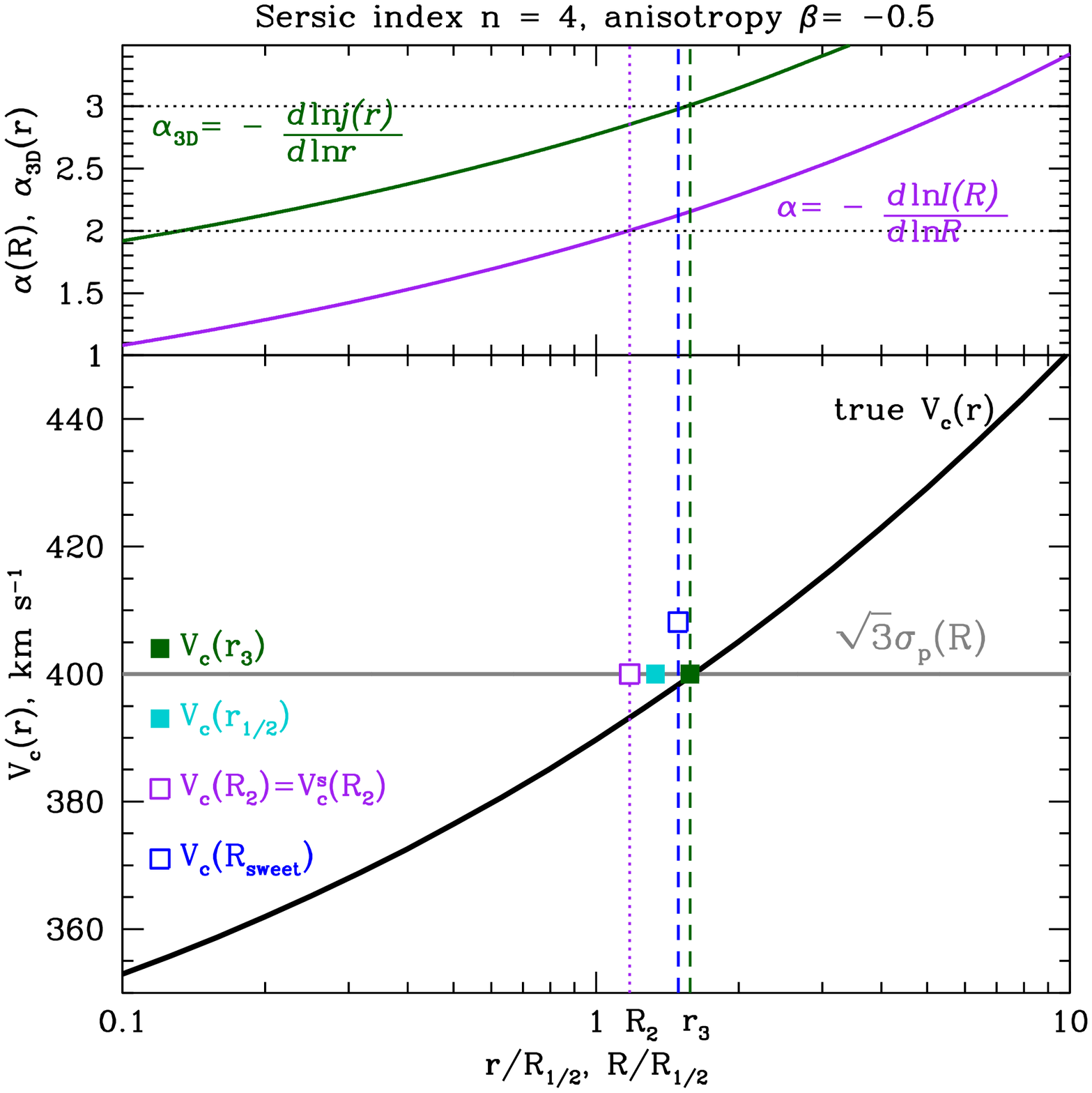}
\caption{Circular speed estimates for an `ideal' model galaxy with flat $V_c(r)$ (left side) or flat $\sigma_p(R)$ (right side). The surface brightness is described by the S\'{e}rsic profile (S\'{e}rsic index $n=4$), and the anisotropy parameter is $\beta=-0.5$. The parameters $n$ and $\beta$ are chosen rather arbitrarily for demonstration purposes only. The upper panel shows the log-slopes of the 3D luminosity density $ \alpha_{3D}$ (in dark green) and of the surface brightness $\alpha$ (in purple). The projected velocity dispersion, the true circular speed profile as well as simple $V_c$-estimates are shown in the lower panel. The symbols in the lower panels are the same as in Figure~\ref{fig:demo}. $r_3$, $R_{\rm sweet}$ and $R_2$ are marked as dark green dashed, blue dashed and purple dotted lines.
\label{fig:ideal}
}
\end{figure*}

First we apply the estimators described in Section \ref{sec:mass_app} and Table~\ref{tab:description} to two `ideal' galaxies, which meet all the assumptions used to derive the local and global formulae: dynamical equilibrium, spherical symmetry, no streaming motions, constant anisotropy, flat $V_c(r)$ for the Churazov et al. formula or constant $\sigma_p(R)$ for the Wolf et al. estimator. We call these two models the `ideal local' and `ideal global' galaxies, respectively. Figure~\ref{fig:ideal} shows typical profiles, derived for an $n=4$ S\'{e}rsic model with constant anisotropy  $\beta=-0.5$, along with the derived local and global $V_c$-estimates. The open blue square shows the local estimate at $ R_{\rm sweet}$ (L1), the open purple square  at 
$ R_{2}$ (L2) and the open purple star is the simplified version of the local estimator (L3). Simple global estimates are shown by filled symbols:  the dark green square is for the global $V_c$-estimate at $r_3$ (G1), the filled cyan square  at $r_{1/2}\approx \frac{4}{3}R_{1/2}$ (G2/G3). 
As expected both estimators 
work well when applied to 
`ideal' galaxies which meet all the corresponding assumptions. In the ideal local case, the global Wolf et. al 
estimators slightly underestimate the true circular speed with typical deviation of $\approx -3\%$. Likewise, in the ideal global case, the local estimators tend to overestimate the true $V_c$ by $\approx 3 \%$ corresponding to mass errors of $\mp 6\%$.

\subsubsection{Grid of analytical models}
\label{subsubsec:model}

\begin{figure*}
\plotone{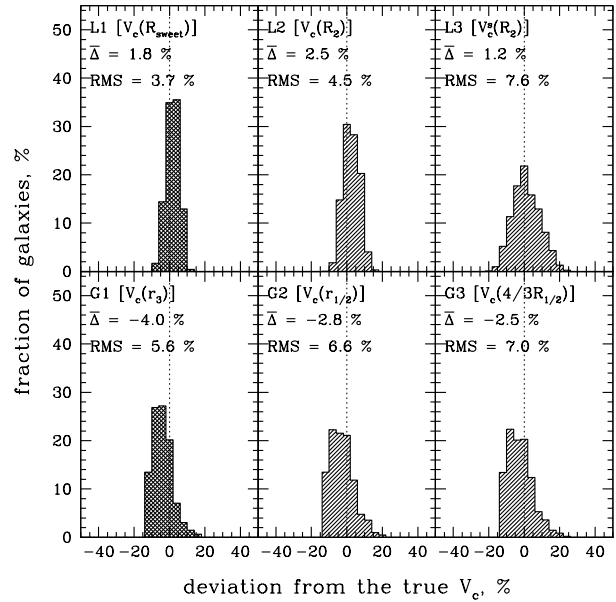}
\caption{The histograms of the deviations of simple $V_c$-estimates from the true value $V_c^{true}$ for the local (upper row) and the  global estimators (lower row) for model spherical galaxies (Section \ref{subsubsec:model}). The analytical models are described by S\'{e}rsic surface brightness profile, anisotropy increasing slowly with radius, and circular speed profile that is similar in shape to observed circular velocity curves. Deviations are calculated as $\Delta= (V_c-V_c^{true}) / V_c^{true}$, where the estimated $V_c$ and true $V_c^{true}$ are taken at the same radius. The RMS-scatter is calculated relative to $\overline{\Delta}$. 
\label{fig:model}
}
\end{figure*}

We explore $\sim30 000$ analytical models, described by the S\'{e}rsic surface brightness profile with index $2 < n < 20$, anisotropy profile increasing slowly with radius and circular speed characteristic for (i) dark-matter dominated dwarf spheroidal galaxies ($V_c$ growing with radius; $\simeq 50\%$ of the models) and (ii) massive elliptical galaxies ($V_c$ roughly flat or decreasing slowly with radius; $\simeq 50\%$ of the models). However, we do not aim to explore the whole parameter space: the idea is to understand how sensitive the estimators are to the assumption of a flat $V_c(r)$ or $\sigma_p(R)$ and to varying anisotropy. The resulting histograms for the local (upper row)  and  global (lower row) estimators are shown in Figure~\ref{fig:model}. The RMS scatter for the global estimators is almost twice  as large as for the local one.

\begin{figure*}
\plottwo{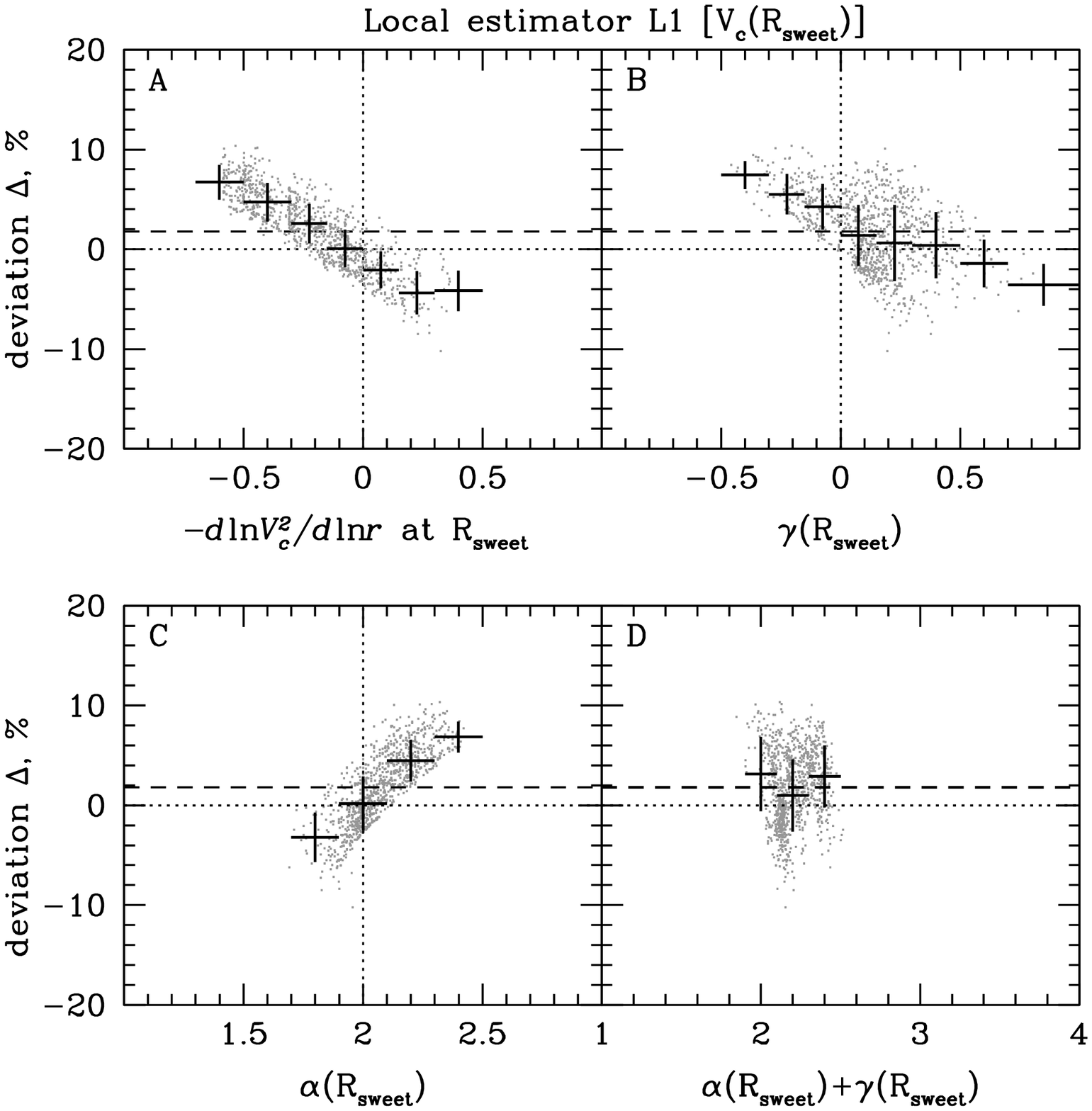}{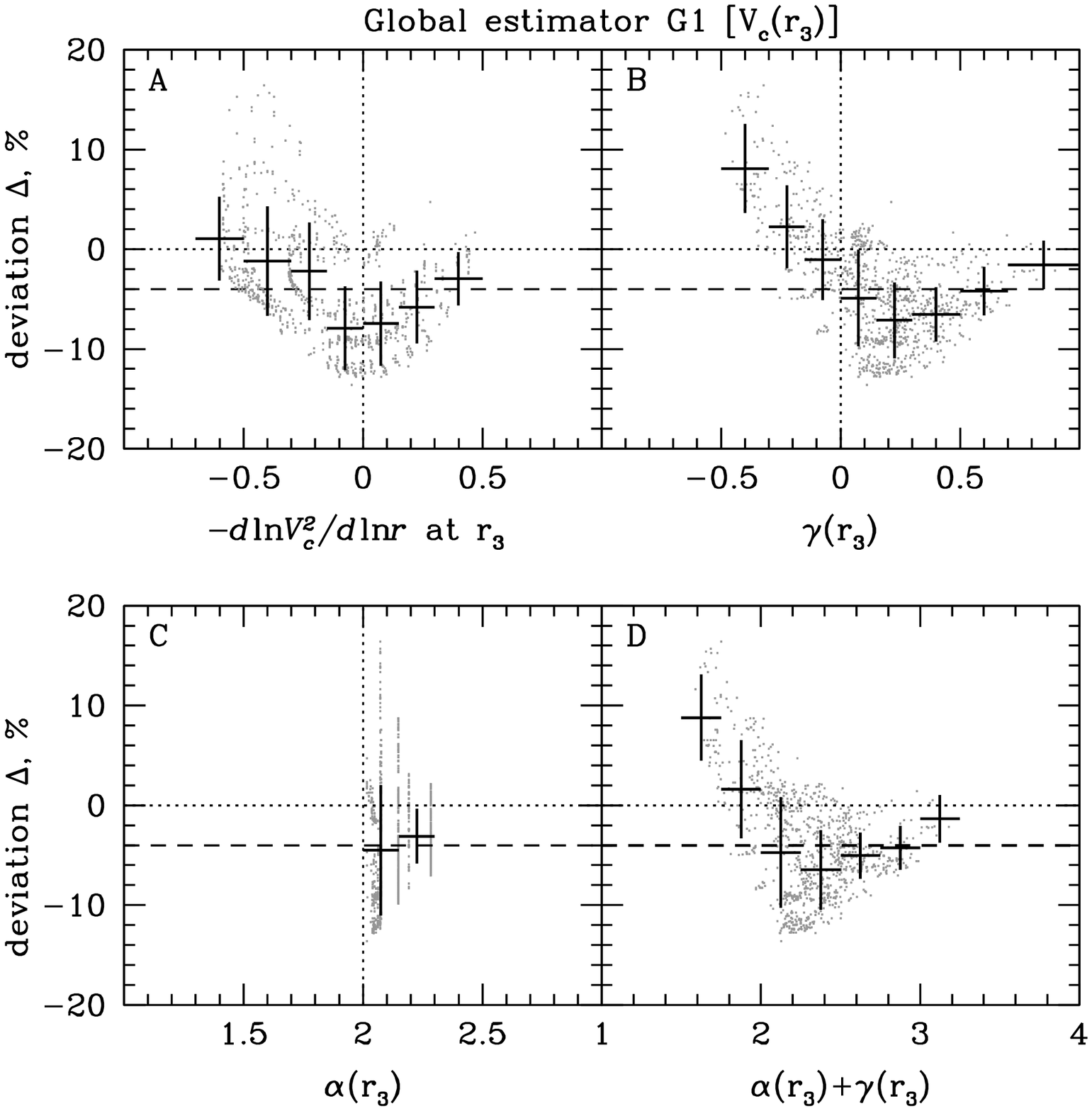}
\plottwos{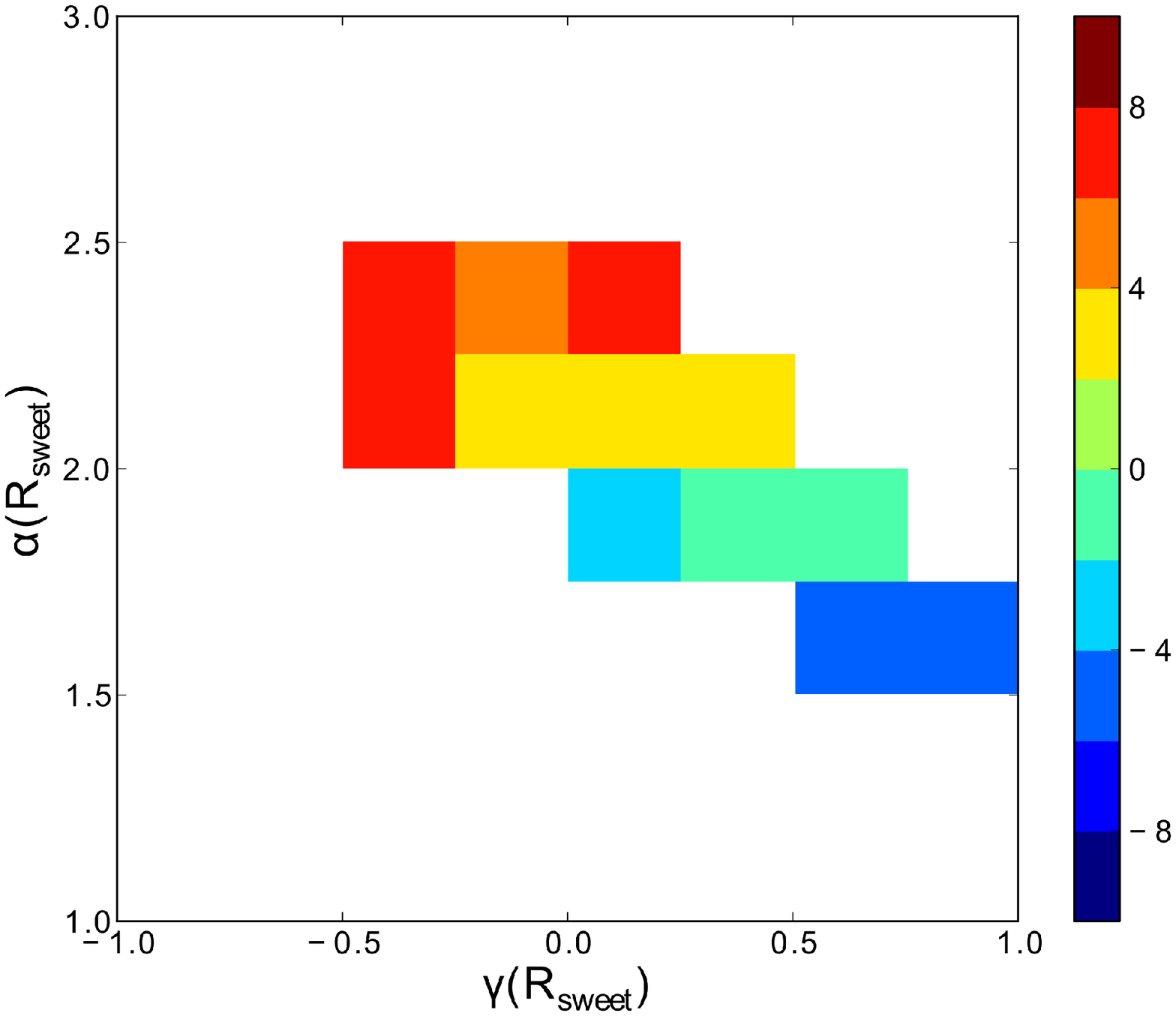}{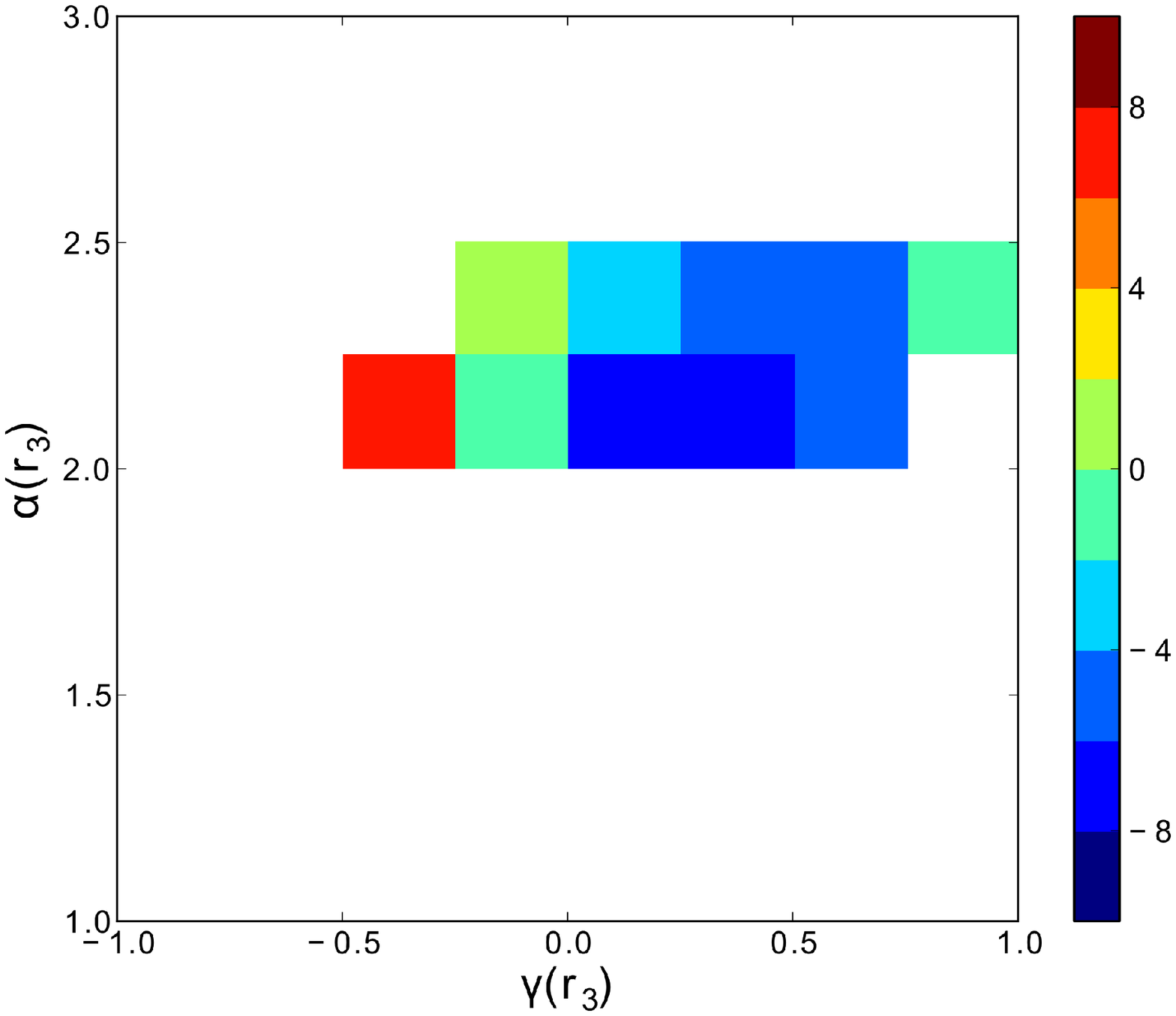}
\caption{Dependence of the error in the $V_c$ estimate on properties of the true $V_c(r)$ and the observable $I(R)$ and $\sigma_p(R)$ profiles for model spherical galaxies (Section \ref{subsubsec:model}). Panels~A: Deviation $\Delta$ of the estimated $V_c$ from the true one as a function of the log-slope of the true circular speed $-d\ln  V^2_c / d\ln r$ taken at the characteristic radius ($ R_{\rm sweet}$ for the local estimator L1 and $r_3$ for the global estimator G1). Panels~B: $\Delta$ as a function of the log-slope of the projected velocity dispersion $ \gamma = -d\ln  \sigma_p^2 / d\ln R$. Panels~C: $\Delta$ as a function of the log-slope of the surface brightness profile $ \alpha = -d\ln I / d\ln R$. Panels~D: $\Delta$ as a function of $\alpha+\gamma$.
The horizontal dashed line shows the deviation averaged over the whole sample of analytical models. The crosses show the average deviation in a chosen bin. Only 1000 randomly chosen realizations are shown for clarity.
Lower row: Color-coded deviation $\Delta$ as a function of $\alpha$ and $\gamma$. Left panels show the local estimator L1 and right panels show the global estimator G1.
\label{fig:corr_model}
}
\end{figure*}

As the  Churazov et al. derivation assumes an logarithmic gravitational potential (i.e., flat $ V_c(r)$) it is important to  test how the accuracy of the estimator depends on the slope of the circular speed profile. Indeed, we find for the local estimator L1 that there is a clear correlation between the fractional deviation $\Delta$ of the estimated circular speed from the true one and the logarithmic slope of the true $V_c(r)$ at $R_{\rm sweet}$: $ \Delta \approx k \times \left( -d\ln  V_c^{2} / d\ln r\right)$ (Figure~\ref{fig:corr_model}, panel~A).  
To understand this trend, let us consider for simplicity pure circular orbits.
If the circular speed increases (decreases) with 
radius, then stellar velocities at $r>R$ are larger than 
at $r=R$, resulting in an increase (decrease) of $\sigma_p$ compared to our reference case of a spatially constant $V_c$. This, in turn, results in a systematic
overestimation (underestimation) of the circular speed. 
The details of the mean correlation between $ \Delta$ and $  -d\ln  V^2_c / d\ln r$ depend on the sampling of the parameter space. However the main trend for the local estimator seems to be rather universal. For growing/decreasing $V_c$ near $R_{sweet}$ (which is close to $R_2$) the method tends to overestimate/underestimate the true value of $V_c$ by a factor of $ \approx 1+0.1 \left( -d\ln  V_c^2 / d\ln r\right)$. For flat $V_c$ near $R_{sweet}$ 
the local estimator is largely unbiased when averaged over the parameter space covered by our grid of analytical models.  
When deviations are plotted against the logarithmic slope of the projected velocity dispersion  $ \gamma = - d\ln  \sigma_p^2 / d\ln R$ at $ R_{sweet}$ (Figure~\ref{fig:corr_model}, panel B), a similar pattern is observed. If $ \sigma_p$ grows with radius in the vicinity of $ R_{\rm sweet}$, then the local $V_c$-estimate  overestimates the true circular speed. For flat or moderately falling observed velocity dispersion profiles the local estimator seems to recover an almost unbiased $V_c$-estimate. 
The observed trend in $ \Delta(\gamma)$ is partly compensated by the opposite trend in $ \Delta(\alpha)$ (Figure~\ref{fig:corr_model}, panel C) to make the coefficient $ \sqrt{1+\alpha(R_{sweet})+\gamma(R_{sweet})}$ relating $\sigma_p$ to the circular speed (first equation in eq.~(\ref{eq:main})) be close to $\sqrt{3}$. The deviation $\Delta$  is almost independent of $ \alpha+\gamma$ (panel D of Figure~\ref{fig:corr_model}) for the local estimator, and the spread in $ (\alpha+\gamma)$ at $ R_{sweet}$ is quite small ($ 1.9 \lesssim \alpha+\gamma \lesssim 2.5$).  
For the local estimators L2 and L3 correlations between $\Delta$ and $  -d\ln  V^2_c / d\ln r$ are quantitatively the same as for L1. 

The global estimator shows a more complex dependence on the slope of the circular speed radial profile (Figure~\ref{fig:corr_model}, panel A), giving a noticeable negative bias for a flat circular speed (for flat $ V_c(r)$ the observed velocity dispersion might vary significantly with radius).  As expected, the Wolf et al. formula works best for roughly flat line-of-sight velocity dispersion profiles. The global $V_c$-estimate appears to be too large for $ \sigma_p$ rapidly increasing with radius ($ \gamma \lesssim 0.3$). Large negative deviations are present for models with $ \sigma_p(R)$ showing a bump, i.e., when $ \gamma(R)$ changes sign. Similar dependencies are observed for the global estimators G2 and G3.

Figure~\ref{fig:corr_model} indicates that the local estimator should be applied with caution to systems with increasing velocity dispersion profiles and/or to systems that are described by growing circular speed profiles in the vicinity of $R_2$. As the global estimator relies on global properties of the galaxies, it works well if the velocity dispersion does not change significantly with radius over the whole extent of the system. Roughly speaking, the  Wolf et al. formula is appropriate for dwarf spheroidal galaxies (see \citealt{Kowalczyk.et.al.2013} who have tested the global estimator on a sample of simulated dSph) and for a subset of elliptical galaxies with approximately flat velocity dispersion profiles, while the local estimator works for elliptical galaxies in general. A similar conclusion is reached when analyzing subsamples of analytical models with $V_c(r)$ resembling that of (i) dwarf spheroidal galaxies and (ii) elliptical galaxies.  

It should also be noted that for large S\'{e}rsic indices ($ n>8-10$), typical for massive ellipticals sitting at the centers of groups or clusters, the log-slope of the surface brightness $ \alpha$ is close to $2$ over a wide 
range of radii, and in this radial range the true circular speed is well described by the isotropic one $V^{\rm iso}_c$ (eq.~(\ref{eq:main})).

\subsection{Simulated galaxies}
\label{subsec:sim}

From spherical models we now turn to a sample of `zoom-in' cosmological simulations of individual galaxies (\citealt{Oser.et.al.2010, Oser.et.al.2012}), which span a wide range in mass, $ 7 \times 10^{11} M_{\odot} h^{-1} < M_{vir} < 2.7 \times 10^{13} M_{\odot} h^{-1}$, $h=0.72$, where $M_{vir}$ is the present day virial halo mass. Those simulated galaxies have properties very similar to observed nearby early-type galaxies as studied by the ATLAS$^{3d}$ project (\citealt{Naab.et.al.2013}).  The typical effective radii of these galaxies are $ \sim 2-15$ kpc. We define the effective radius to be $R_{1/2}$, that is, the radius of a circle which contains half of the {\it total} stellar mass of a galaxy without introducing any cut-off (in contrast to  \cite{Oser.et.al.2010, Oser.et.al.2012}  who define $R_{1/2}$ as a radius which encloses half of the projected stellar mass within 10\% of the virial radius). 
The typical axis ratios {\it q} (ellipticity $\epsilon = 1-q$), calculated as the square root of eigenvalues of the diagonalised inertia tensor within the effective radius, are $\sim 0.5-1$. The anisotropy of the simulated galaxies is mildly tangential or close to isotropic at the center and becomes radially anisotropic with $\beta \sim 0.2-0.4$ at large radii \citep{Wu.et.al.2014}. The surface brightness profiles of simulated galaxies are well discribed by  cored S\'ersic models with typical S\'ersic indices $n \gtrsim 10$   \citep{Wu.et.al.2014}.

The local estimators have been tested in \cite{Lyskova.et.al.2012} and we follow the same analysis procedure. Briefly, we first exclude satellites from the galaxy image and calculate the radial profiles $I(R)$, $\sigma_p(R)$ and the true circular speed $ V_c^{true}(r) = \sqrt{GM(<r)/r}$ in a set of logarithmic concentric annuli/shells around the halo center.  In practice, such averaging over concentric annuli could be applied to 2D kinematic maps derived from integral field unit observations. At $R_{\rm sweet}$ we take $ V_c^{\rm iso}$ as an estimate of the circular speed and calculate the deviation from the true value at this radius $ \Delta=\left(V_c^{\rm iso}-V_c^{\rm true}\right)/V_c^{\rm true}$. 

We apply the estimators to a subsample of massive ($\sigma_p(R_{1/2})>150$ $\kms$) simulated galaxies and exclude merging ($\sim 3\%$ of the total) and oblate galaxies seen almost along the rotation axis ($\sim 15\%$ of the total).

\begin{figure*}
\plotone{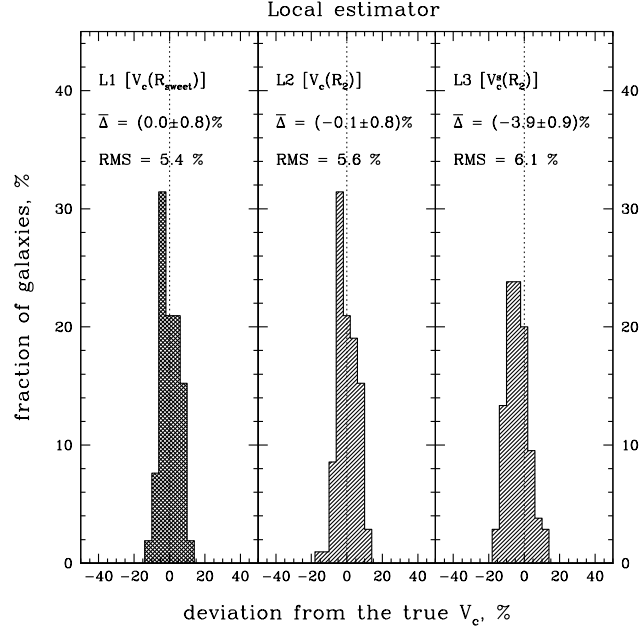}
\caption{The histograms of the deviations of the local $V_c$-estimators from the true circular speed for simulated galaxies (Section~\ref{subsec:sim}). $ V_c(R_{\rm sweet})$ (L1) and $ V_c(R_{2})$ (L2) are calculated from eq.~(\ref{eq:main}) at $ R_{\rm sweet}$ (radius where $V^{\rm iso}_c \approx V^{\rm circ}_c \approx V^{\rm rad}_c$) and $ R_2$ (where $ I \propto R^{-2} $) respectively. $ V_c^{s}(R_{2})$ (L3) comes from eq.~(\ref{eq:agd_simple}).
\label{fig:histo_EC}
}
\end{figure*}

Figure~\ref{fig:histo_EC} shows the fraction of galaxies  versus the deviation $ \Delta=\left(V_c-V_c^{\rm true}\right)/V_c^{\rm true}$ for the local estimators L1, L2, and L3. The left and right panels have already been presented in  \cite{Lyskova.et.al.2012} (their Figure~8). Note that L1 and L2 estimators show very similar results: the average deviation over the sample is close to zero and the RMS scatter $\approx 5-6 \%$. So the radius $R_2$, which is uniquely determined from the slope of the surface brightness profile, can be used instead of $R_{\rm sweet}$, which depends also on the log-slope of the projected velocity dispersion profile $-d\ln  \sigma_p^2 / d\ln R$ and the second derivative $d^2\ln[I\sigma_p^2] / d  (\ln R)^2$. 

\begin{figure*}
\plotone{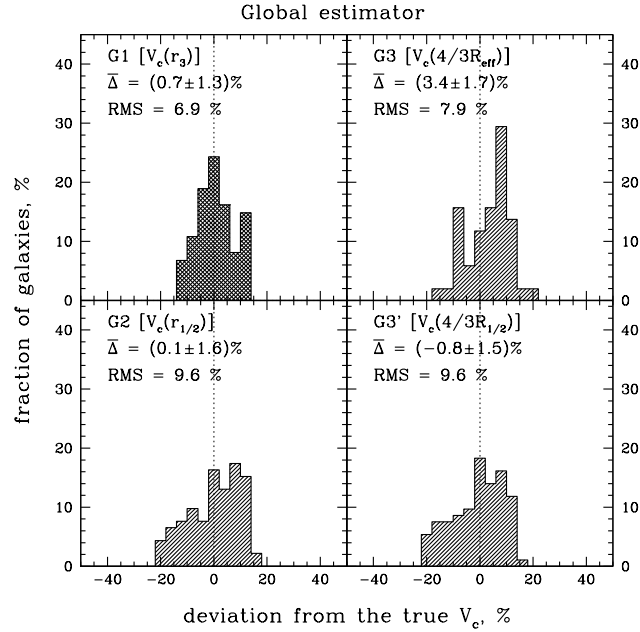}
\caption{ The histograms of the deviations of the  global $V_c$-estimators from the true circular speed for simulated galaxies (Section~\ref{subsec:sim}). Deviations are calculated at the following  characteristic radii: (i) $r_3$ (upper left panel), where $j(R) \propto r^{-3}$, (ii) a factor $\frac{4}{3}$ times the `observational' effective radius resulting from the S\'{e}rsic fit to the surface brightness over $[3r_{soft};0.1R_{vir}]$ (upper right panel), (iii) 3D half-light radius $r_{1/2}$ (lower left panel) and (iv) a factor $\frac{4}{3}$ times the projected half-light radius $R_{1/2}$ (lower right panel). 
\label{fig:histo_W}
}
\end{figure*}

To test the robustness of the Wolf et al. estimator on the sample of simulated galaxies we need to compute 
(i) the luminosity-weighted projected velocity dispersion $ \langle \sigma_p^2 \rangle$ over some radial range $[R_{min};R_{max}]$, 
(ii) the characteristic radii $r_3$, $r_{1/2}$ and $\frac{4}{3}R_{1/2}$ where the anisotropy is expected to play a minimal role in the circular speed determination. 
The 3D half-light radius $r_{1/2}$ and  projected half-light radius $R_{1/2}$ are computed as the radius of the sphere/circle which contains half of the 3D/projected stellar mass, respectively. Since in real observations information on the total light is often not available or difficult to estimate, we also determine an `observational' effective radius $R_{\rm eff}$ coming from a S\'{e}rsic fit $I(R) \propto \exp\left( -b_n(R/R_{\rm eff})^{1/n}\right)$ to the observed surface brightness over the same radial range $[R_{min};R_{max}]$ used for the calculation of $ \langle \sigma_p^2 \rangle$. We choose $R_{min}=3r_{soft}$, where $r_{soft} \approx 400$ $h^{-1}$ $\rm pc$  is the softening length of the simulations and $3r_{soft}$ is the maximum radius where profiles could be affected by the softening. Finally, $R_{max}=0.1R_{vir}$, where $R_{vir}$ is the halo virial radius ($\equiv R_{200}$, the radius where the spherical overdensity drops below 200 times the critical density of the Universe).

Figure~\ref{fig:histo_W} shows the perfomance of the Wolf et al. estimator at different radii: 
(G1) $r_3$ (upper left panel), where $\alpha_{3D}=- d\ln j(r) / d\ln r =3$, 
(G2) 3D half-light radius $r_{1/2}$ (lower left panel),
(G3) $\frac{4}{3} \times$ `observational' effective radius (upper right panel) and
(G3$'$) $\frac{4}{3} \times$ projected half-light radius $R_{1/2}$ (lower right panel). 
While the global estimates $V_c(r_3)$ and $V_c(r_{1/2})$  are almost unbiased (when averaged over the sample of simulated galaxies), the average deviation of the global $V_c\left(\frac{4}{3}R_{\rm eff}\right)$ is $\approx 3-4\%$ high. The RMS scatter for all cases is equal to $\approx 7-10\%$, i.e., significantly larger than for the local estimators. As mentioned above, observed profiles at $ R<R_{min}=3r_{soft}$ could be affected by the softening, and for the analysis of the Wolf et al. estimator we consider only those simulated galaxies for which the characteristic radius is larger than $ 2R_{min}=6r_{soft}$. This selection criterion 
effectively keeps only the most massive galaxies with roughly logarithmic gravitational potential (see Section 3.2 in \citealt{Lyskova.et.al.2012}) in the sample. For these galaxies the virial theorem states that $V_c$ at any radius is well approximated by $3\langle\sigma_p^2 \rangle$ and the exact value of the characteristic radius is not important.

If we vary  $R_{max}$ and $R'_{max}$ within reasonable limits, the values of $R_{\rm eff}$ and $\langle \sigma_p^2 \rangle$ for individual galaxies do change, but not dramatically, and the average deviation $\Delta$ remains practically the same ($|\overline{\Delta}| \lesssim 3-5\%$) with RMS scatter  $\simeq 8-10 \%$. 

\begin{figure*}
\plottwo{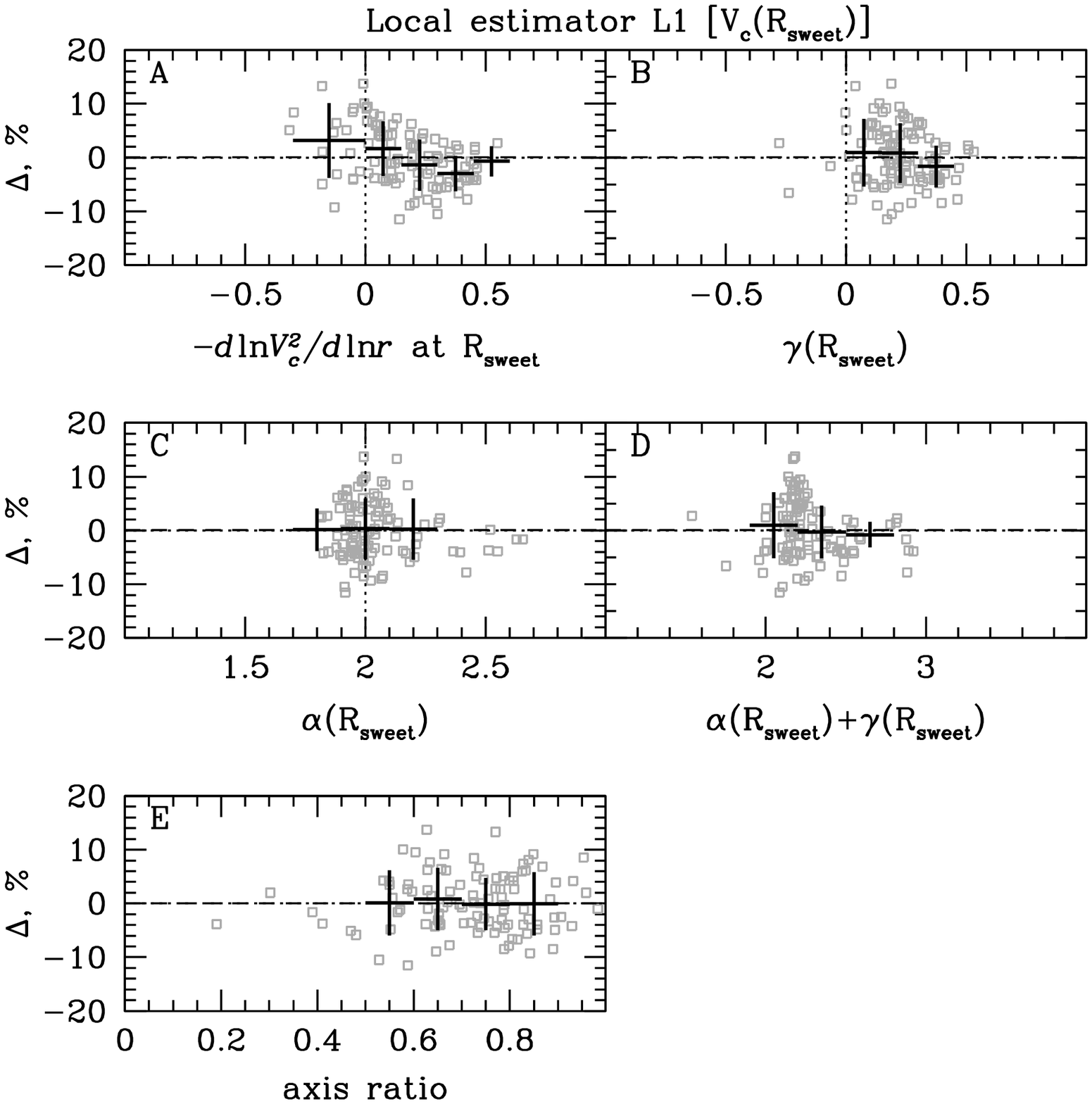}{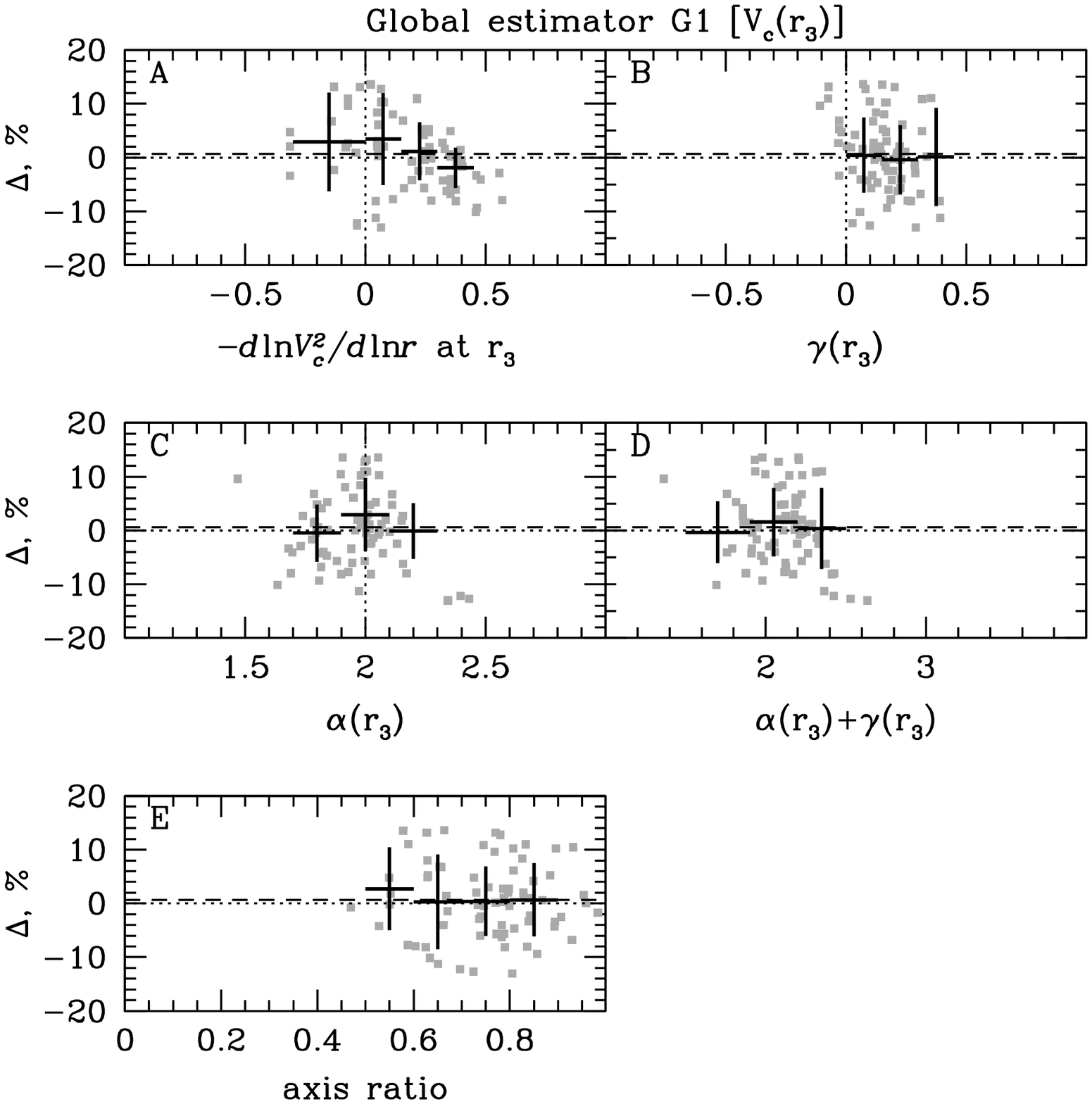}
\caption{Observed correlations for simulated galaxies (Section~\ref{subsec:sim}). Panels~A: Deviation $\Delta$ as a function of the log-slope of the true circular speed $ -d\ln  V^2_c / d\ln r$ taken at a characteristic radius ($R_{\rm sweet}$ for the local estimator L1 and $r_3$ for the global estimator G1). Panels~B: $\Delta$ as a function of the log-slope of the projected velocity dispersion $\gamma = -d\ln  \sigma_p^2 / d\ln R$. Panels~C: $\Delta$ as a function of the log-slope of the surface brightness profile $\alpha = - d\ln I / d\ln R$. Panels~D: $\Delta$ as a function of $\alpha+\gamma$. Panels~E:  $\Delta$ as a function of axis ratio. The horizontal dashed line shows the deviation averaged over the whole sample of simulated galaxies. The crosses show the average deviation in a chosen bin. Left panels show the local estimator L1 and right panels show the global estimator G1.
\label{fig:corr_sim}
}
\end{figure*}

Tests on model analytical galaxies (Section \ref{subsubsec:model}) show that for the local estimator the deviation $\Delta$ correlates with the logarithmic slope of the true circular speed. For the rising $V_c(r)$ the local approach tends to overestimate the true circular velocity. For the sample of simulated galaxies we also observe such a trend in the local estimator L1 (Figure~\ref{fig:corr_sim}, Panels~A), but not as strong as for the model spherical galaxies. For the local estimators L2 and L3 the trend is similar. In the probed region of $\gamma$ the deviation $\Delta$ is fairly flat (Figure~\ref{fig:corr_sim}, Panel D) as is for the spherical analytical models (Figure~\ref{fig:corr_model}, Panel D).

No clear correlation between simple local and global estimates and the logarithmic slopes of the projected velocity dispersion $\gamma$ or the log-slope of the surface brightness $\alpha$ or the axis ratio is found.

\begin{figure*}
\plottwos{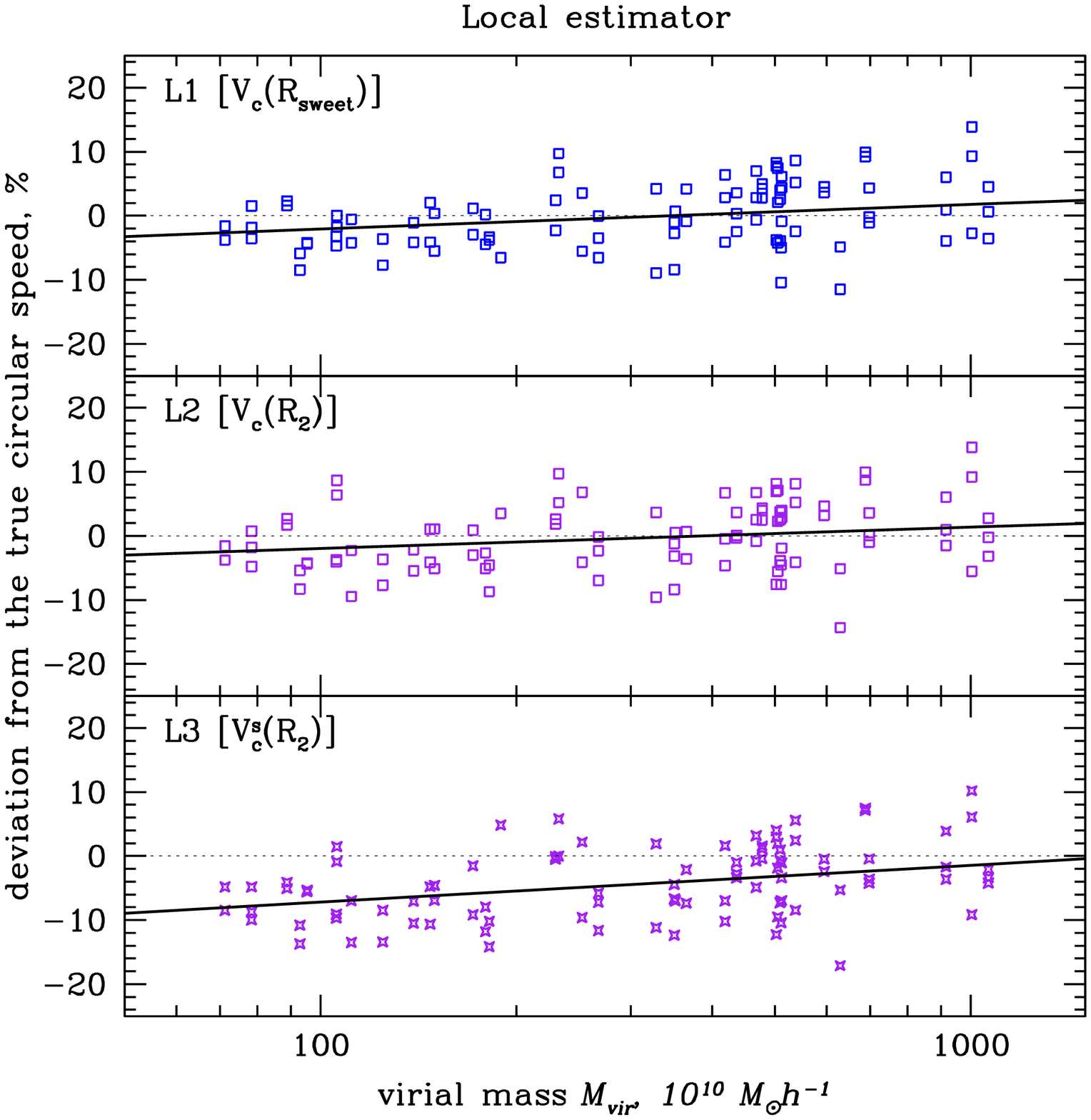}{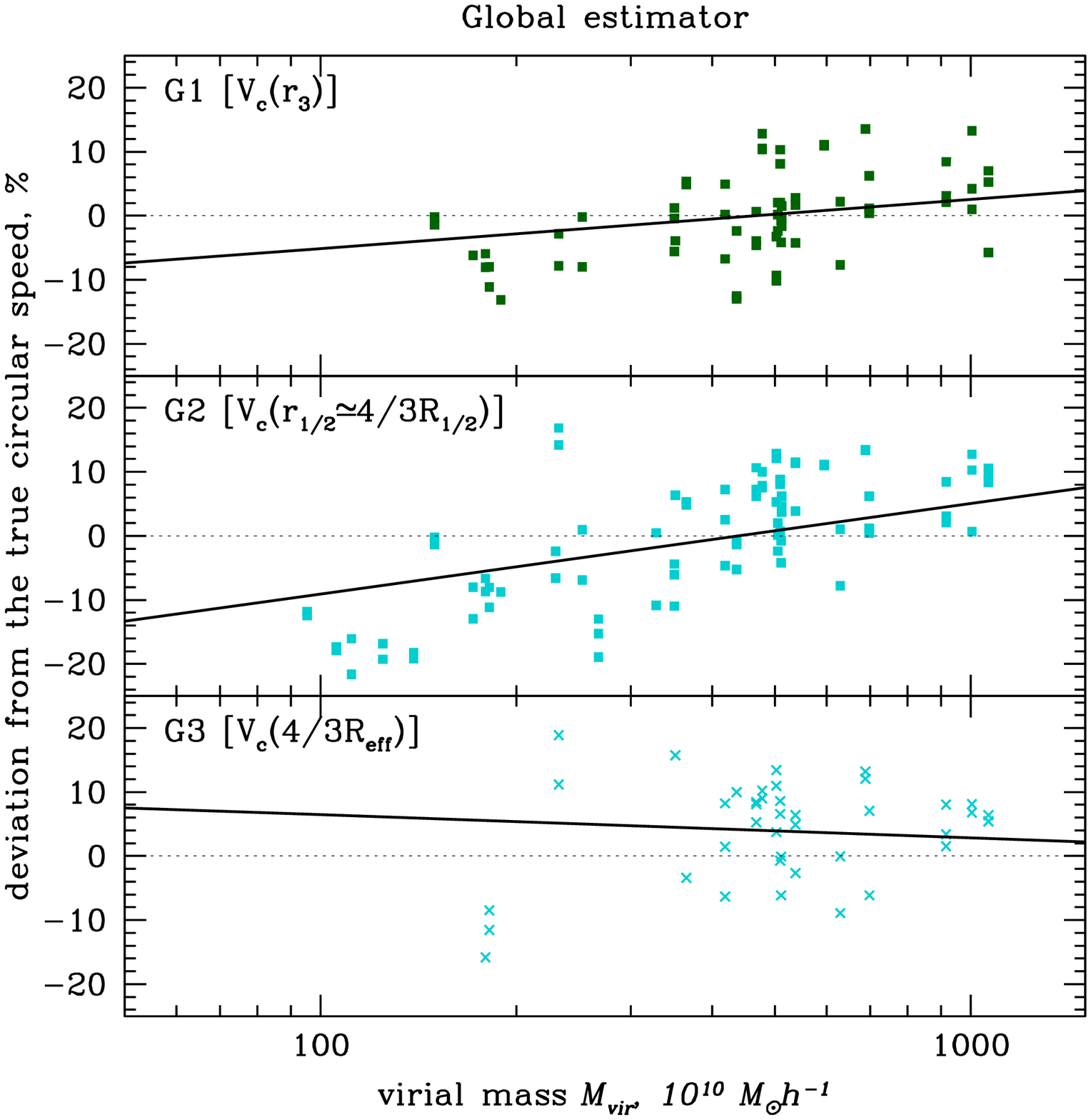}
\caption{Deviation of the estimated circular speed from the true one as a function of the virial halo mass (extracted from simulations). The local formula (left side) recovers the true circular speed for a wide range of virial masses almost equally well. The global approach works better for the most massive galaxies in the sample ($M_{vir} \gtrsim 3\cdot 10^{12}M_{\odot}h^{-1}$), than for the less massive galaxies. One can notice that the global estimator at different characteristic radii probes a different range of virial masses. When using the global estimator at $R_{char} = \left\{r_3, r_{1/2}, \frac{4}{3}R_{\rm eff}\right\}$ we retain in the sample only the galaxies which have $R_{char}>2R_{min}=6r_{soft}$. As a result the three panels for $R_{char} = \left\{r_3, r_{1/2}, \frac{4}{3}R_{\rm eff} \right\}$ probe different subsamples at the low-mass end. In addition, the characteristic radii $r_3$ and $r_{1/2}$, which are expected to be close to each other for commonly used analytical stellar light 
distributions (e.g., for S\'ersic models) sometimes differ by a factor $\sim 2-3$ in the simulated galaxies.}
\label{fig:trends}
\end{figure*}

In Figure~\ref{fig:trends} we show the deviation of simple estimates from the true circular speed as a function of the virial halo mass $M_{vir}$. Dots represent the deviations for individual galaxies and lines are the least-square linear fits of $\disp \Delta[\%]=a\cdot \lg \left(\frac{M_{vir}}{10^{10}M_{\odot}h^{-1}}\right)+b$. Note that for the local estimators L1 and L2 and the global estimator G1 the trends with the virial mass are quite weak. The Wolf et al. formula has been applied only to simulated galaxies with the characteristic radius $R_{char} \ge 2R_{min}=6r_{soft}$, so the probed range of masses is different for different characteristic radii.

\section{Comparison of simple mass estimators with a state-of-the-art analysis}
\label{sec:art}

We now proceed to test the simple mass estimators on real early-type galaxies which have high-quality resolved stellar kinematical data and which have been studied in detail, using  Schwarzschild modeling. The number of such galaxies is constantly increasing, but the extent of the kinematic data for the majority of them is limited to $\approx R_{\rm eff}$ (an `observational' effective radius based on the S\'ersic fit) only. For a successful mass determination with the simple estimators described here, however, it is desirable to have spatially resolved kinematics at least out to $\approx 1.2-1.5 R_{\rm eff}$. As our target sample we therefore use early-type galaxies from a Coma cluster survey, modeled using the Schwarzschild method by \cite{Thomas.et.al.2007b}, and, in addition, the giant elliptical galaxy M87 from the Virgo cluster modeled by \cite{Murphy.et.al.2011}. For our purposes we choose only those galaxies from the Thomas et al. sample where kinematic measurements are available out to $\gtrsim 1.5 R_{\rm eff}$ and the galaxy is slowly rotating in the sense that $\sigma_p(R_{\rm eff})>V_{rot}(R_{\rm eff})$, where $\sigma_p$ and $V_{rot}$ are the projected velocity dispersion and 
rotation velocity, both measured along the major axis. 

\begin{table*}
\centering
\caption{\label{tab:galaxies} Sample of real elliptical galaxies analyzed using the Schwarzschild modeling (7 Coma galaxies from \protect\cite{Thomas.et.al.2007b} and M87 (NGC4486) from \citealt{Murphy.et.al.2011}). The effective radii and ellipticities of the Coma galaxies are taken from \protect\cite{Thomas.et.al.2007b}, the effective radius and the ellipticity of M87 comes from \protect\cite{Kormendy.et.al.2009}.} 
\begin{tabular}{cccc}
\hline
 name & other common names & $R_{\rm eff}$, arcsec & ellipticity $\epsilon$ at $R_{\rm eff}$ \\
\hline
NGC 4957 & GMP 0144 & 18.4 & 0.256\\ 
NGC 4952 & GMP 0282 & 14.1 & 0.315\\
NGC 4908 & GMP 2417 & 7.1 & 0.322 \\
NGC 4869 & GMP 3510 & 7.6 & 0.112\\
IC 3947  & GMP 3958 & 3.3 & 0.323\\ 
NGC 4827 & GMP 5279 & 13.6 & 0.205 \\
NGC 4807 & GMP 5975 & 6.7 & 0.170\\
NGC 4486 & M87      & 194.41 & 0.218\\ 
\hline
\end{tabular}
\end{table*}

\begin{figure*}
\plottwos{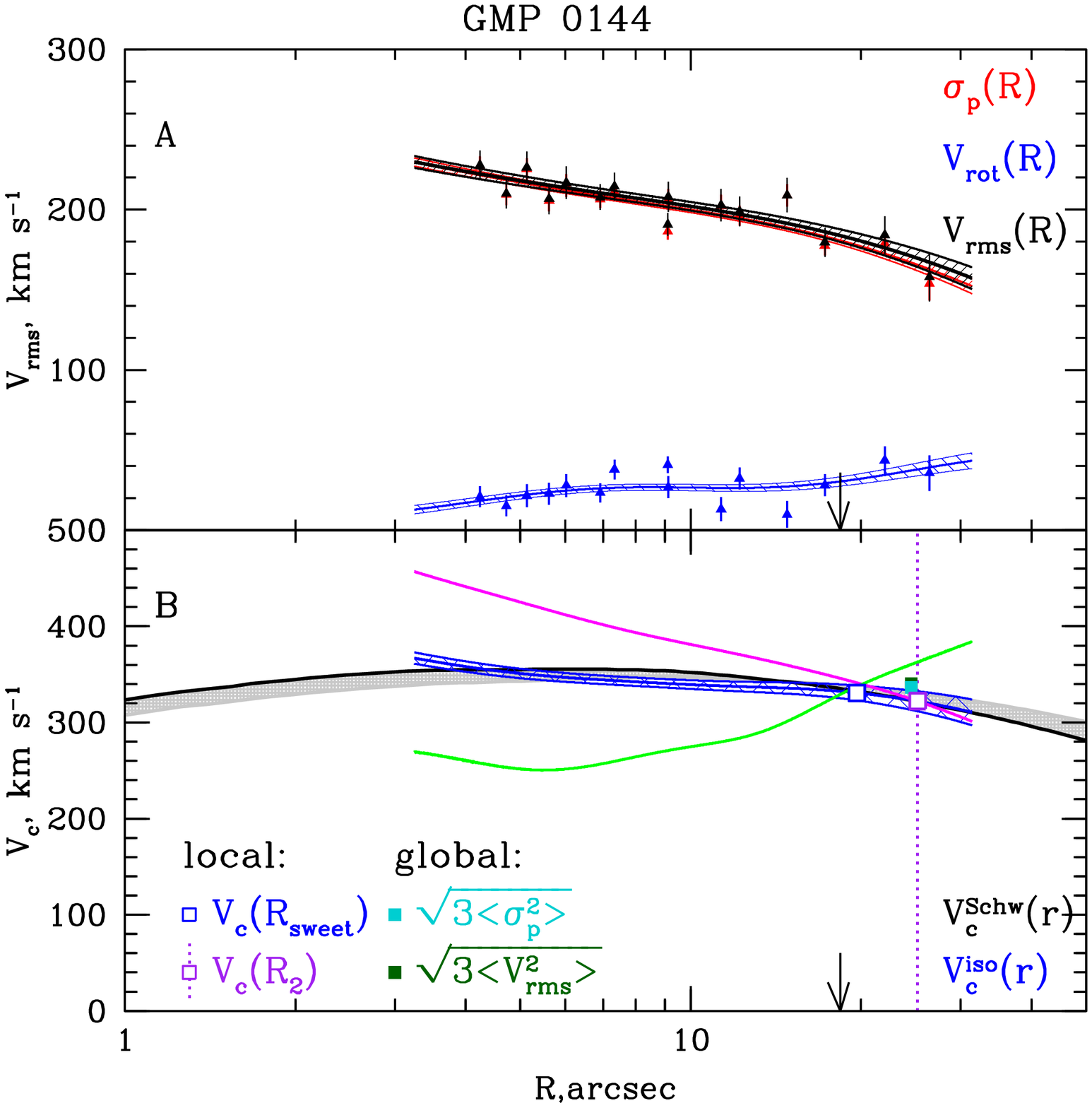}{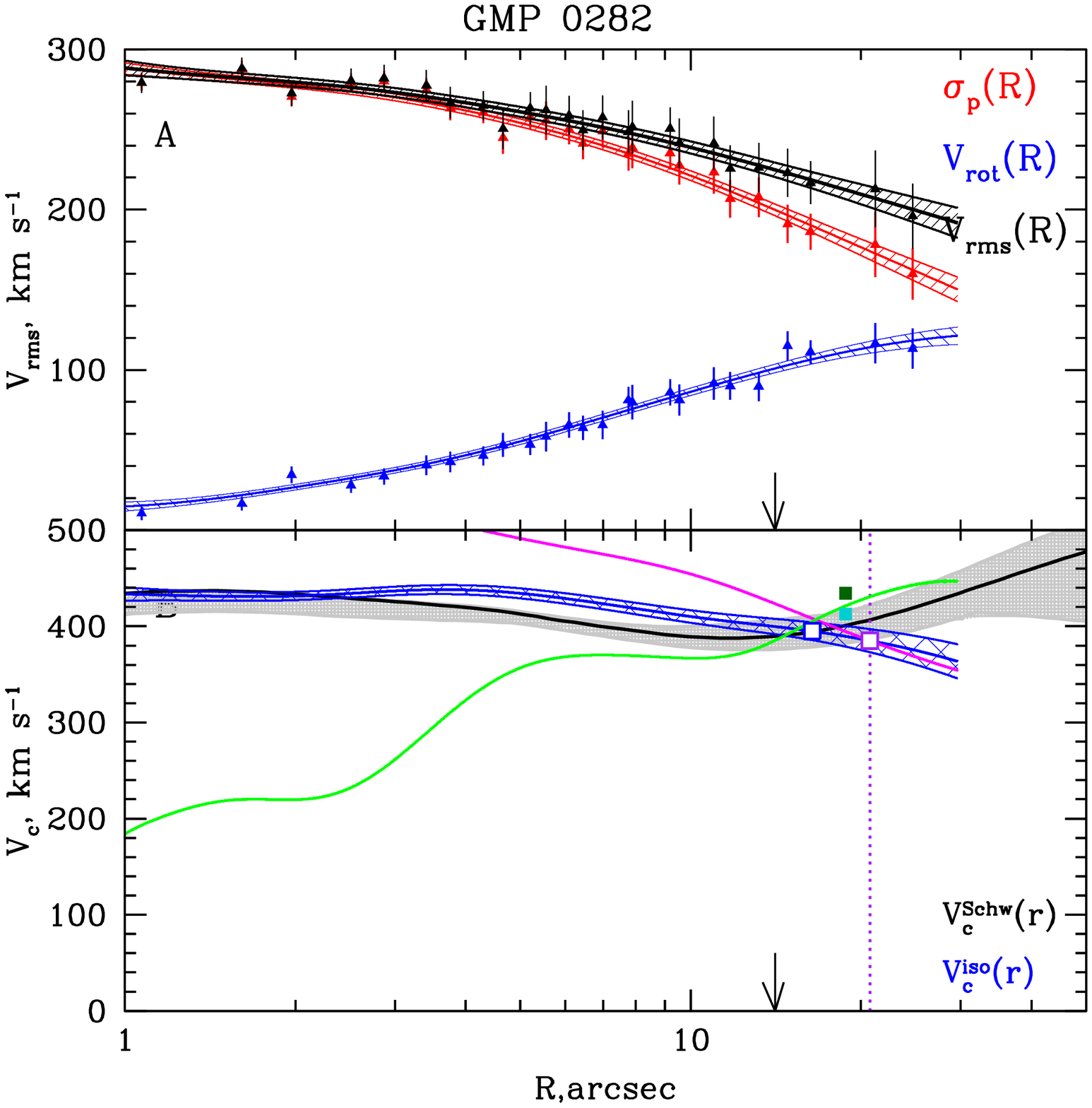}
\plottwos{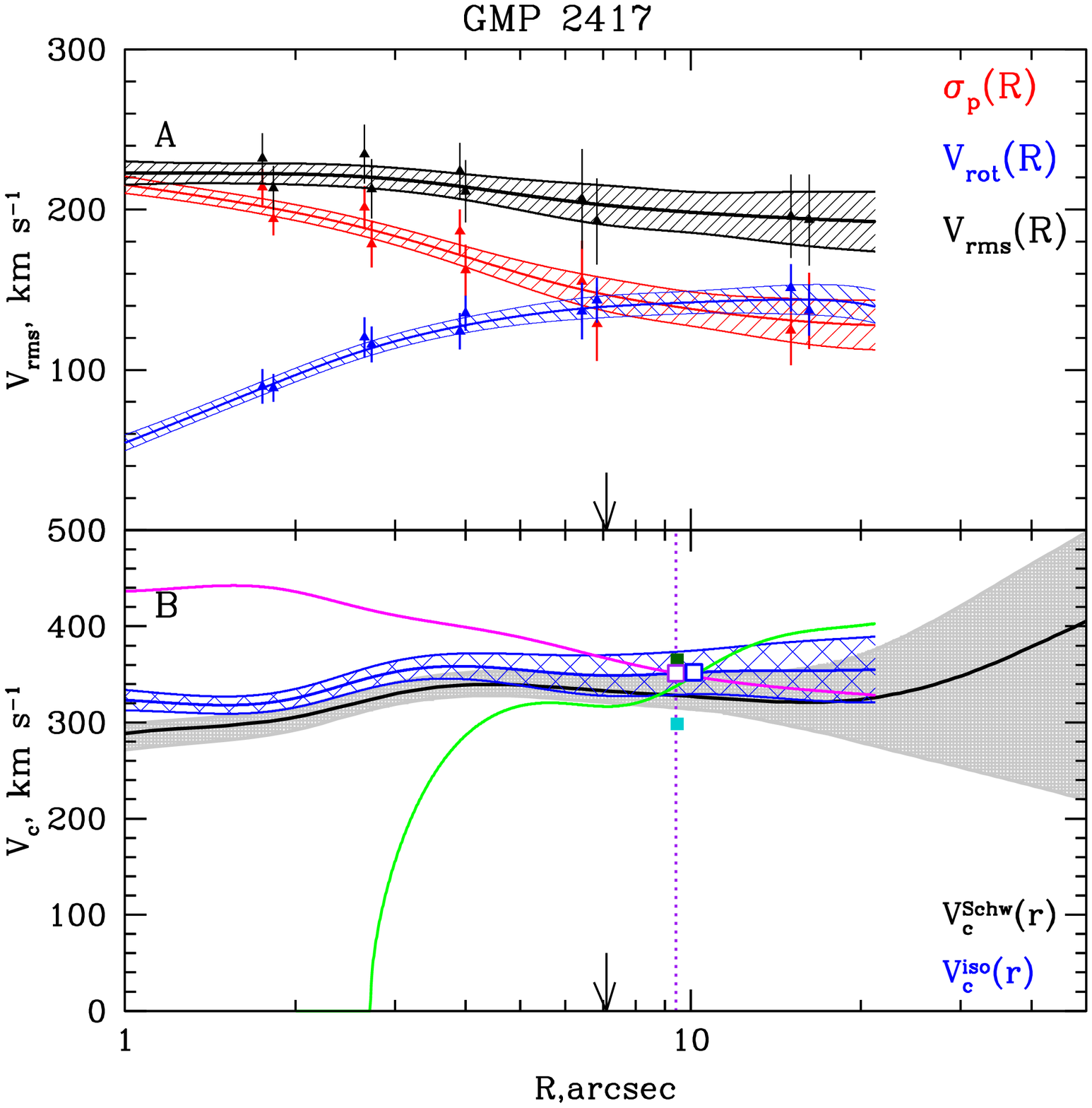}{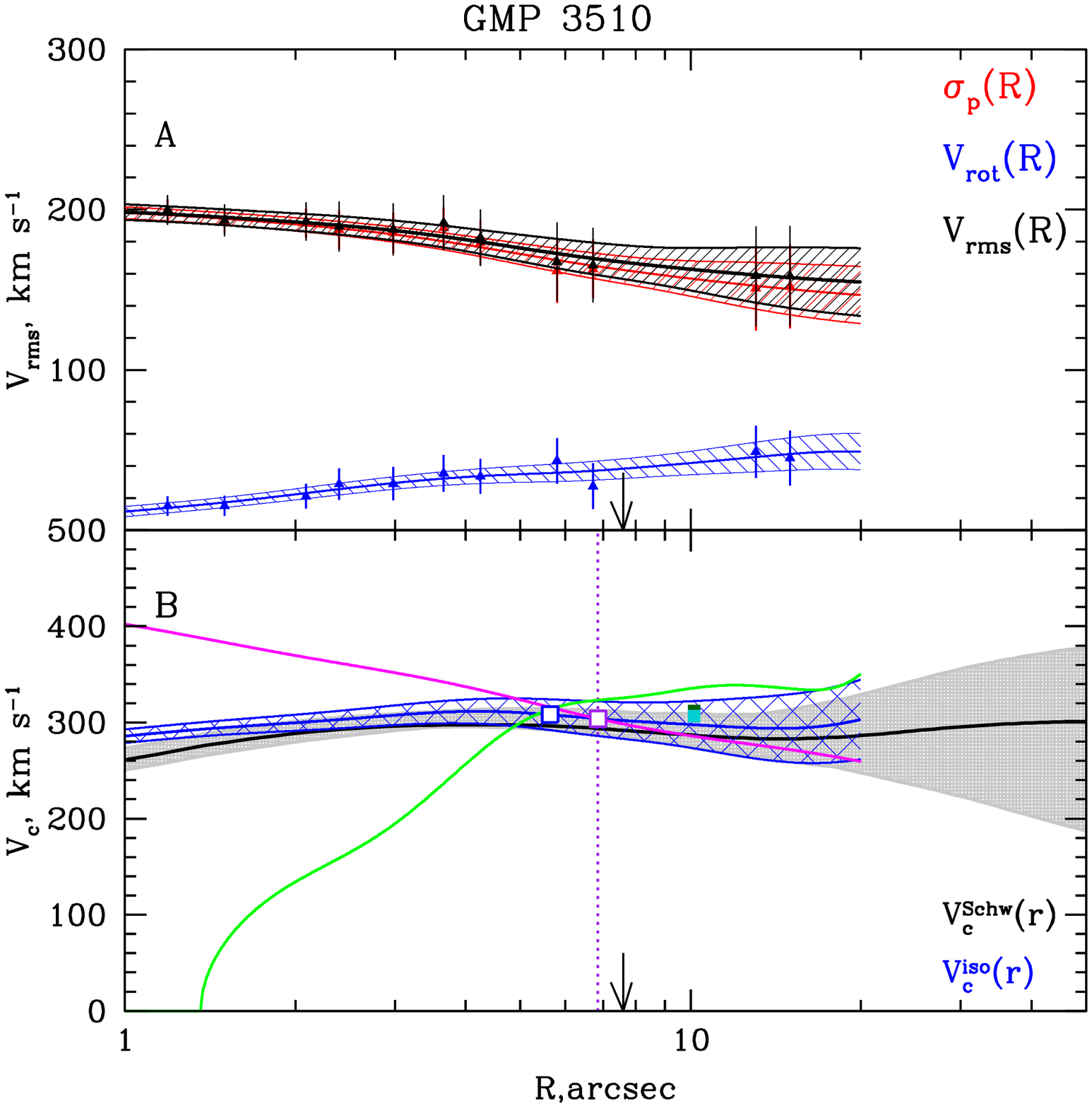}
\caption{Comparison of simple $V_c$-estimates with the circular speed ($V_c^{Schw}$) coming from the dynamical modeling (Section \ref{sec:art}). Panel A shows data on the projected velocity dispersion $ \sigma_p$ and rotation velocity $ V_{rot}$, both measured along the major axis, and RMS velocity ($ V_{RMS}=\sqrt{\sigma_p^2+V_{rot}^2}$) as well as interpolated curves used to calculate the logarithmic derivatives. Panel B presents the circular speed resulting from the Schwarzschild modeling (black thick curve) with error bars (grey shaded region), the isotropic circular speed $ V^{\rm iso}_c$ (in blue) calculated via eq.~(\ref{eq:main}) with `observational' error bars $\Sigma_{obs}$ determined from the measurement uncertainties on $V_{RMS}$. The circular speed profiles for pure circular and pure radial orbits are shown as magenta and green curves respectively. Simple $V_c$-estimates are shown as squares of different colors: the blue open square corresponds to the local $V_c$-estimate at $R_{\rm sweet}$ (L1), the purple open square at $R_2$ (L2), global estimates  (calculated as $ \langle \sigma_p^2 \rangle$   and $ \langle V_{RMS}^2 \rangle$) at $\frac{4}{3}R_{\rm eff}$ (G3) are shown as the cyan and dark green filled squares. The arrow and the dotted purple line show the effective radius $R_{\rm eff}$ and $R_2$ respectively.
\label{fig:art1}
}
\end{figure*}

\setcounter{figure}{9}
\begin{figure*}
\plottwos{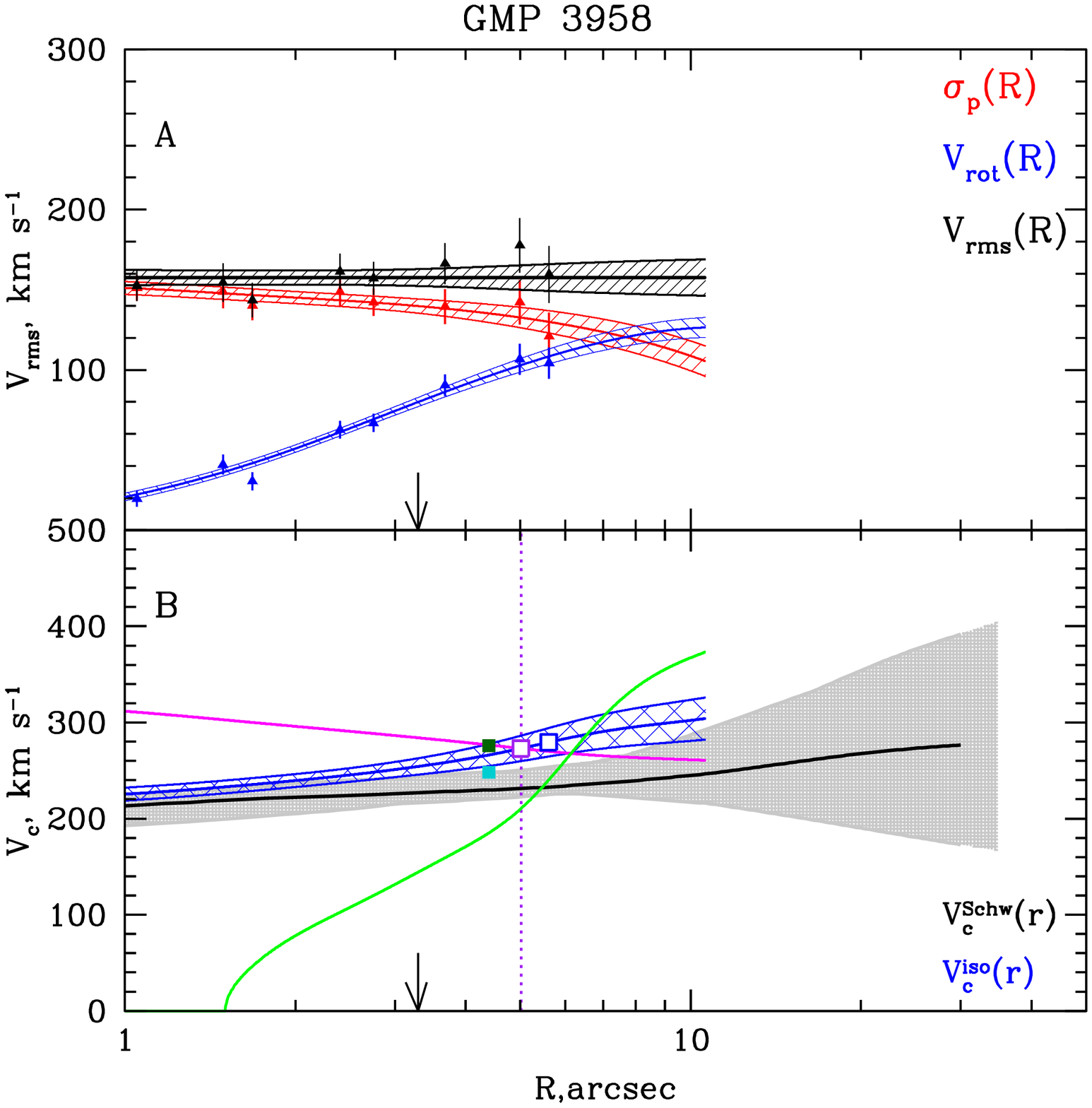}{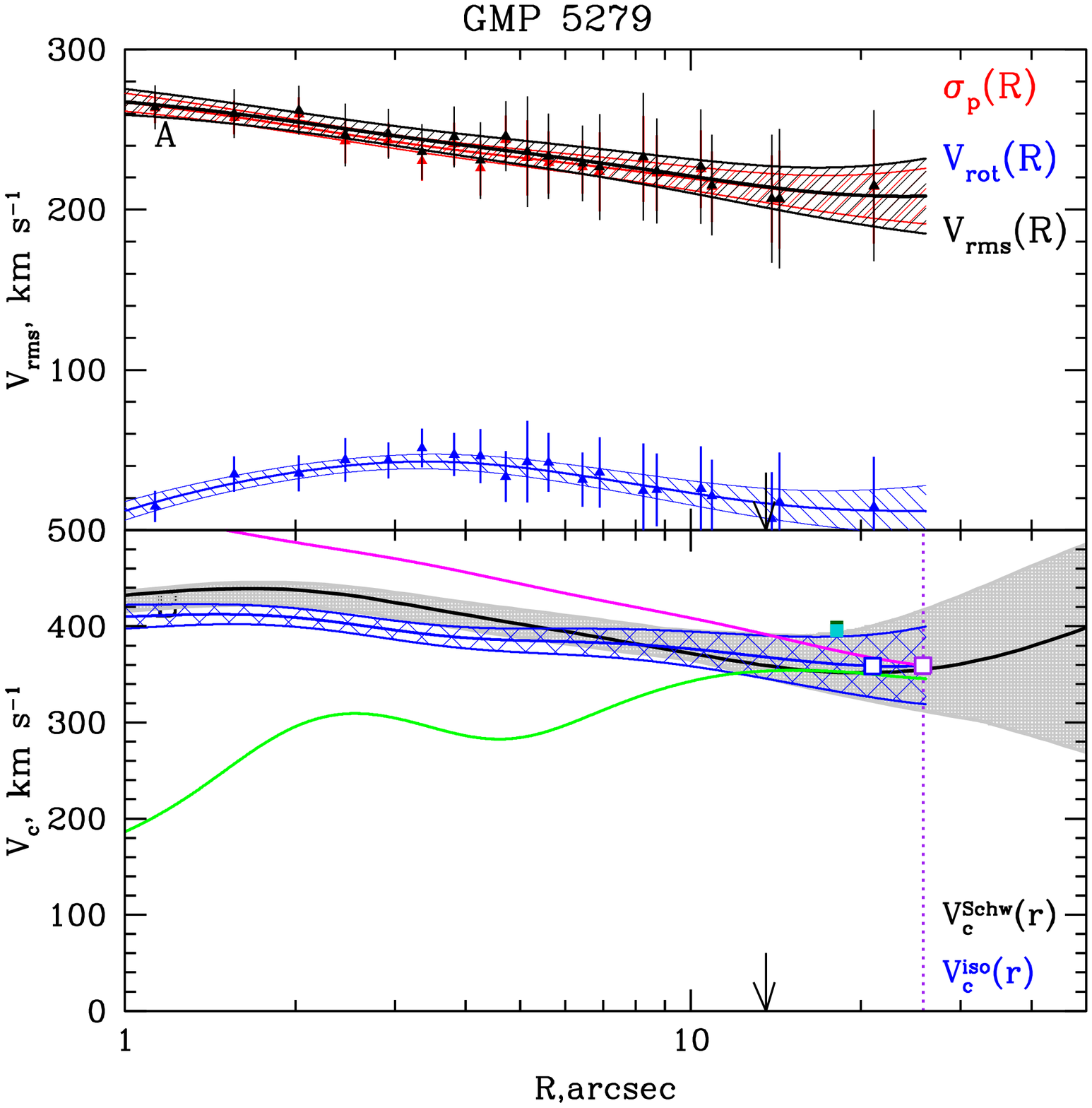}
\plottwos{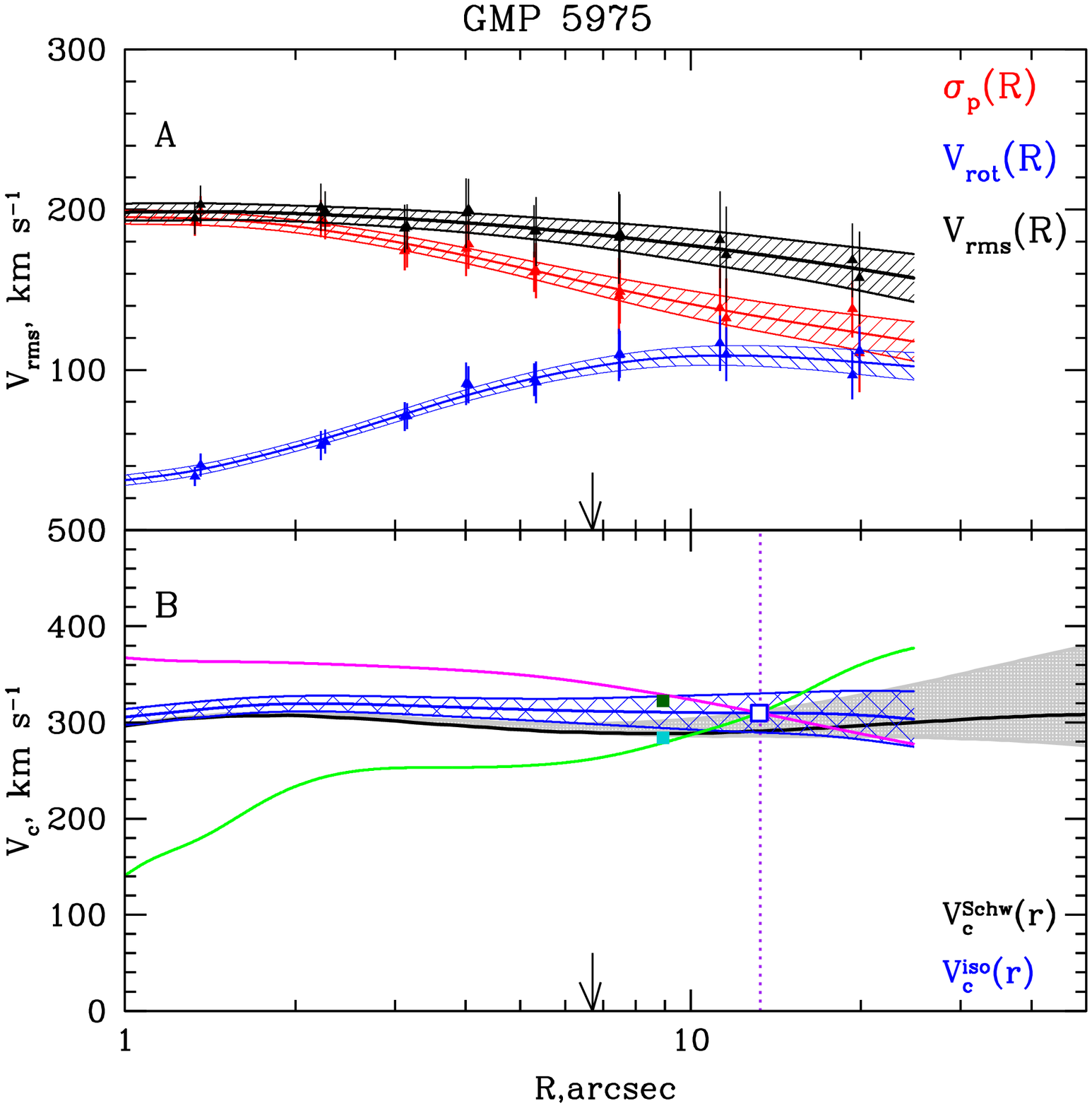}{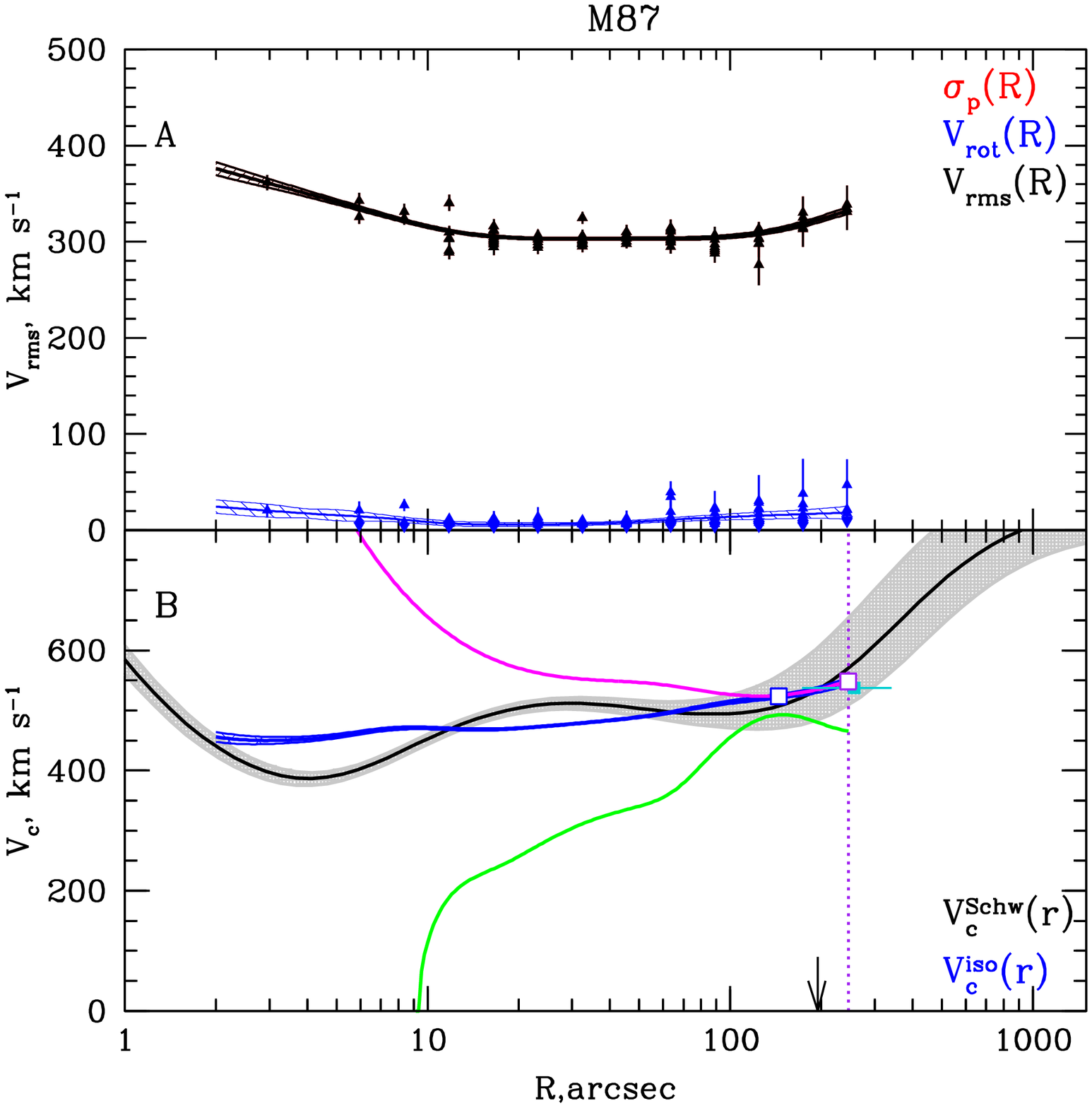}
\caption{(continue)} 
\label{fig:art2}
\end{figure*}

Galaxies of our target sample (7 Coma galaxies + M87) are listed in Table~\ref{tab:galaxies}. We use the surface brightness and the root-mean-square velocity ($V_{RMS}(R)= \sqrt{\sigma_p^2(R)+V_{rot}^2(R)}$) to recover the circular speed  via eq.~(\ref{eq:main}) and  (\ref{eq:w3}). 
For each galaxy we compare simple circular speed estimators  with the best-fit circular velocity. The luminosity-weighted velocity dispersion $ \langle \sigma_p^2 \rangle$ (which is needed for the Wolf et al. estimator) is averaged over the radial range where the observational data are available. As some galaxies show a moderate rotation, we test also a slightly modified version of the Wolf et al. formula where $ \langle \sigma_p^2 \rangle$ is substituted by the average luminosity-weighted RMS velocity $ \langle V_{RMS}^2 \rangle$.

\begin{table*}
\centering
\caption{\label{tab:galaxies2} Simple $V_c$-estimates and $V_c$ from dynamical modeling. The columns are: (1) - the simple $V_c$-estimator; (2) - Deviation of the estimated $V_c$ from $V_c^{Schw}$ resulting from dynamical modeling, averaged over 
the sample of 8 galaxies (7 Coma galaxies + M87); (3) - RMS scatter around the average deviation; (4) - average observational error at the characteristic radius normalised to $V_c^{Schw}$; (5) - average error on dynamically derived $V_c^{Schw}$ at the characteristic radius. Since the bias for the local estimators is mostly driven by a single galaxy IC 3947 (the most compact galaxy in the sample), we exclude this galaxy and provide the results of the analysis of the remaining galaxies in the parentheses.}
\begin{tabular}{lccccc}
\hline
\noalign{\vskip 1mm} 
estimator   & $\disp \bar{\Delta} = \langle V_c/V_c^{Schw}-1 \rangle,\%$ & RMS scatter, \% & $\langle \Sigma_{obs}\rangle/V_c^{Schw}, \%$ & $\langle \Sigma_{Schw} \rangle /V_c^{Schw}, \%$ \\
(1)   & (2) & (3) & (4) & (5) \\
\hline 
\noalign{\vskip 1mm} 
L1 [$\disp  V_c(R_{\rm sweet})$] & 5.2  (3.2) & 6.6 (3.1)& $\pm 5.8$ ($\pm 5.7$)& $^{+8.1}/_{-4.8}$ \hspace{3mm} ($^{+8.0}/_{-4.9}$)\\[0.15cm]

L2 [$\disp V_c(R_{2})$]        &  3.3 (1.2) & 7.5 (4.9) & $\pm 6.2$ ($\pm 6.3$)& $^{+9.6}/_{-6.4}$ \hspace{3mm} ($^{+9.7}/_{-6.7}$)\\[0.15cm]

G3 [$\disp \sqrt{3 \langle \sigma_p^2\rangle}$ at $\frac{4}{3}R_{\rm eff}$] & 2.0 (1.1) & 7.4 (7.6) & $\pm 5.7$ ($\pm 5.8$)&  $^{+8.4}/_{-5.4}$ \hspace{3mm} ($^{+8.4}/_{-5.6}$) \\[0.15cm]

$\disp \sqrt{3 \langle V^2_{RMS}\rangle}$ at $\frac{4}{3}R_{\rm eff}$    & 8.9 (7.3) & 7.8 (6.9) & $\pm 5.7$ ($\pm 5.8$)&  $^{+8.4}/_{-5.4}$ \hspace{3mm} ($^{+8.4}/_{-5.6}$) \\
\hline
\end{tabular}
\end{table*}

The results are presented in Figure~\ref{fig:art1}. Panel A shows the rotational velocity $V_{rot}$ (blue triangles) and projected velocity dispersion $\sigma_p$ (red triangles) measurements along the major axis with errorbars ($\Sigma_{rot}$ and  $\Sigma_{\sigma}$ correspondingly). The RMS velocity $V_{RMS}=\sqrt{\sigma_p^2+V_{rot}^2}$ with observational errors $ \Sigma_{RMS} = \sqrt{\Sigma_{rot}^2+\Sigma_{\sigma}^2}$ is plotted as black triangles. The interpolated curves (the interpolation procedure is decsribed in \citealt{Churazov.et.al.2010}) for the kinematic data are shown as solid lines. Shaded regions show the measurement uncertainties.  Panel B shows the best-fitting circular speed $ V_c^{Schw}$ from the Schwarzschild orbit-superposition modeling (black curve) with 1$\sigma$-uncertainties (grey shaded region); plotted in blue is the isotropic circular speed calculated from eq.~(\ref{eq:main}) where $\sigma_p(R)$ is replaced with $ V_{RMS}(R)$. The blue shaded region reflects the uncertainties associated 
with $V_{RMS}$-measurements. Simple $V_c$-estimates are shown as squares of different colors: the blue open square corresponds to the local $V_c$-estimate at $R_{\rm sweet}$ (L1), the purple open square at $R_2$ (L2), global estimates  (calculated as $ \langle \sigma_p^2 \rangle$   and $ \langle V_{RMS}^2 \rangle$) at $\frac{4}{3}R_{\rm eff}$ (G3) are shown as the cyan and dark green filled squares. The effective radius $R_{\rm eff}$ is shown with an arrow, and the radius $R_2$ as a purple dotted line. As the `true' (coming from the Schwarzschild modeling) circular speed and observed line-of-sight velocity dispersion (or RMS velocity) profiles are close to being flat, both Churazov et al. and Wolf et al. estimators are expected to recover the circular speed reasonably well. Note that galaxies in the chosen subsample do not match perfectly all the requirements for using the simple mass estimators. Namely, most of them are flattened and some  are slowly rotating. Nevertheless, we see that the simple $V_c$-estimates agree well 
with $V_c^{Schw}$, especially for slowly rotating galaxies (GMP 0144, GMP 3510, GMP 5279), where simple estimates (local $V_c$-estimates at $R_{\rm sweet}$ and $R_{2}$) almost coincide with the $V^{Schw}_c(r)$ value. Table~\ref{tab:galaxies2} summarizes the comparison of simple $V_c$-estimates with the circular velocity from the Schwarzschild modeling. The estimators used are listed in Column~(1), the mean deviation $ \bar{\Delta}$ from $V_c^{Schw}$  in Column~(2). RMS scatter around $ \bar{\Delta}$ (in Column~(3)) is calculated as $\disp \sqrt{\frac{\sum_{i=1}^{N}(\Delta_i-\bar{\Delta})^2}{N-1}}$. Column~(4) shows the uncertainty $ \Sigma_{obs}$ associated with measurement errors. For each galaxy $ \Sigma_{obs}$ is calculated as $ \sqrt{1+\alpha+\gamma}\cdot\Sigma_{RMS}$ (eq.~(\ref{eq:main})).  The column (5) lists the average errors on the best-fit circular speed from the Schwarzschild analysis taken at the corresponding characteristic radius. 
For real elliptical galaxies the characteristic radii $r_{1/2}$ and $r_3$ may not be available, and their determination often requires additional assumptions, so in Table~\ref{tab:galaxies2} we provide results for the global estimate at $\frac{4}{3}R_{\rm eff}$ only. Furthermore, elliptical galaxies in the target sample have roughly flat circular speed and projected velocity dispersion profiles, so the performance of the global estimator at all characteristic radii ($r_{3}$, $r_{1/2}$ and $\frac{4}{3}R_{\rm eff}$) is expected to be similar. The average deviation of the global $V_c(r_3)$ 
and $V_c(r_{1/2})$ are $\simeq 2.6 \%$ and $\simeq 1.3 \%$ high, and the RMS deviations are $\simeq 6.0 \%$ and $\simeq 8.5 \% $, correspondingly. Here, the radii $r_3$ and $r_{1/2}$ are obtained from the spherical deprojection of the observed surface brightness profile and its S\'ersic fit, correspondingly.

\begin{figure*}
\plottwo{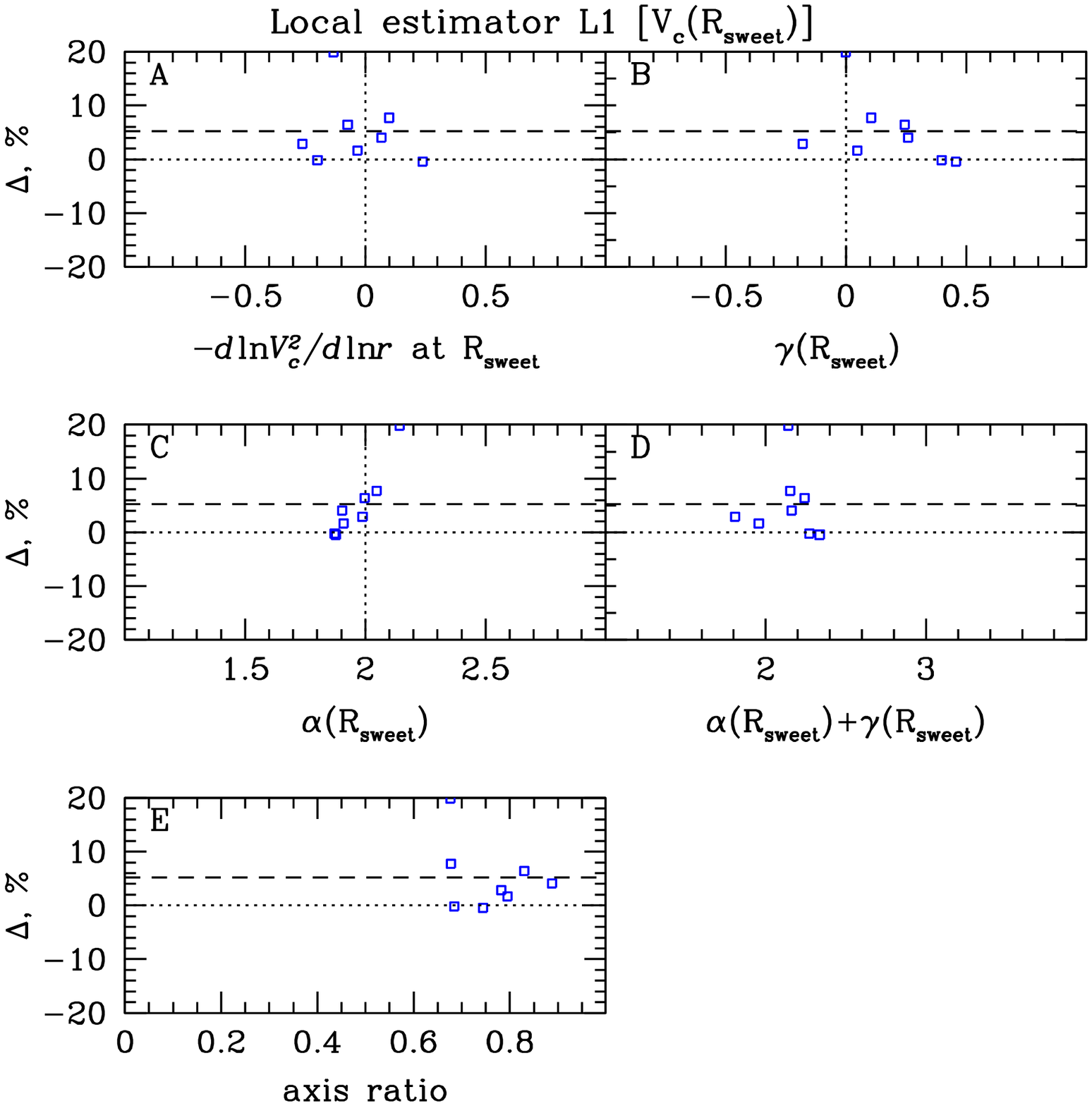}{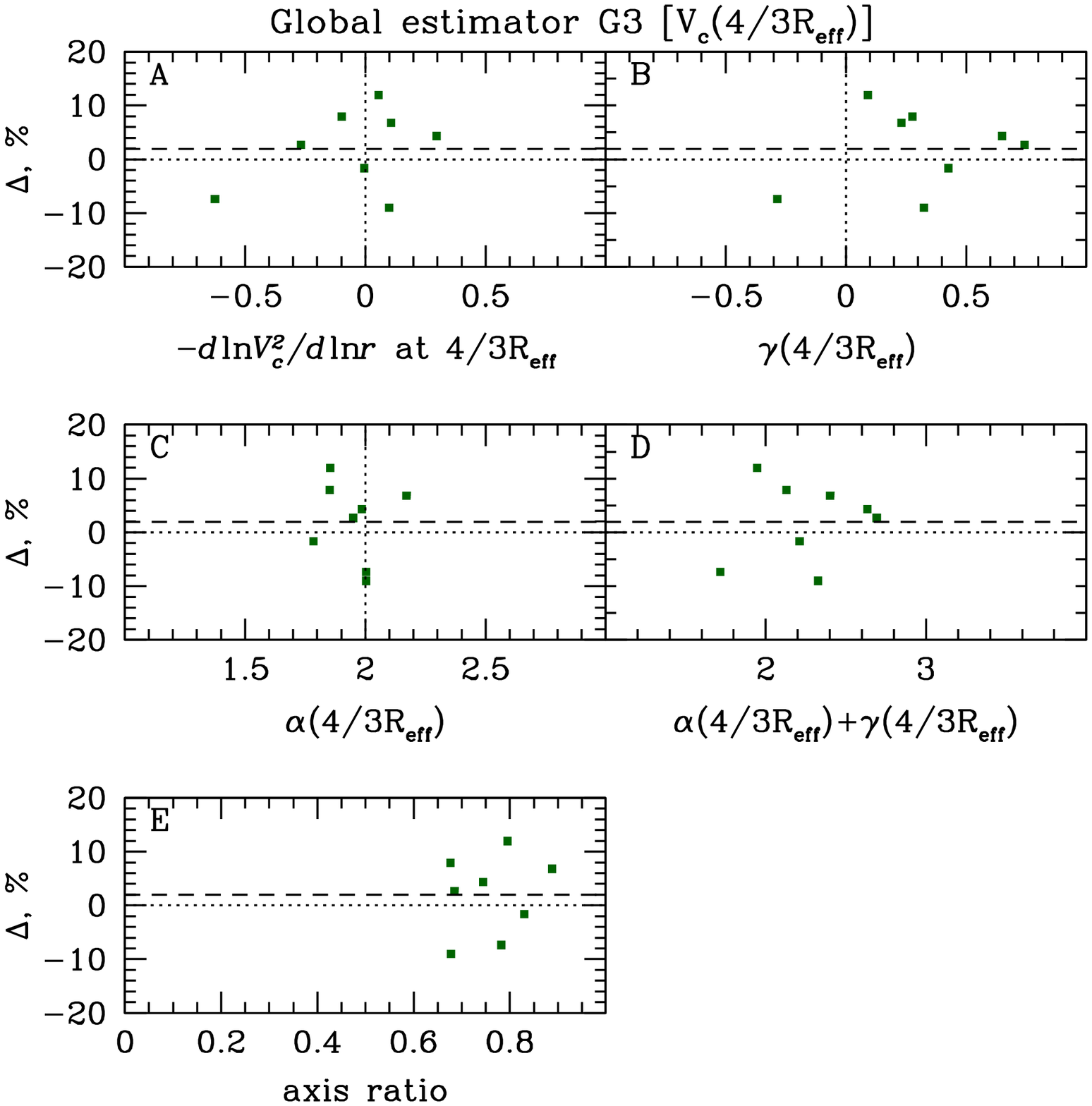}
\caption{Observed correlations for real galaxies (7 Coma galaxies + M87). Panels~A: Deviation $\Delta$ as a function of the log-slope of the circular speed from dynamical modeling taken at a characteristic radius ($R_{\rm sweet}$ for the local estimator L1 and $\frac{4}{3}R_{eff}$ for the global estimator G3). Panel~B: $\Delta$ as a function of the log-slope of the projected velocity dispersion $\gamma = -d\ln  \sigma_p^2 / d\ln R$. Panels~C: $\Delta$ as a function of the log-slope of the surface brightness profile $\alpha = - d\ln I / d\ln R$. Panels~D: $\Delta$ as a function of $\alpha+\gamma$.     Panels~E:  $\Delta$ as a function of axis ratio $1-\epsilon$. The horizontal dashed line shows the deviation averaged over the whole sample of elliptical galaxies. The histogram shows the average deviation in a chosen bin. Left panels show the local estimator L1 and right panels show the global estimator G3.
\label{fig:corr_real}
}
\end{figure*}

Figure~\ref{fig:corr_real} shows the observed correlations for the target sample of real early-type galaxies. Again, there is no clear correlation between simple $V_c$-estimates and axis ratios of galaxies, although the selection criterion on the rotational velocity ($ \sigma_p>V_{rot} $ at the effective radius) leaves only relatively round galaxies.

\begin{figure*}
\plottwos{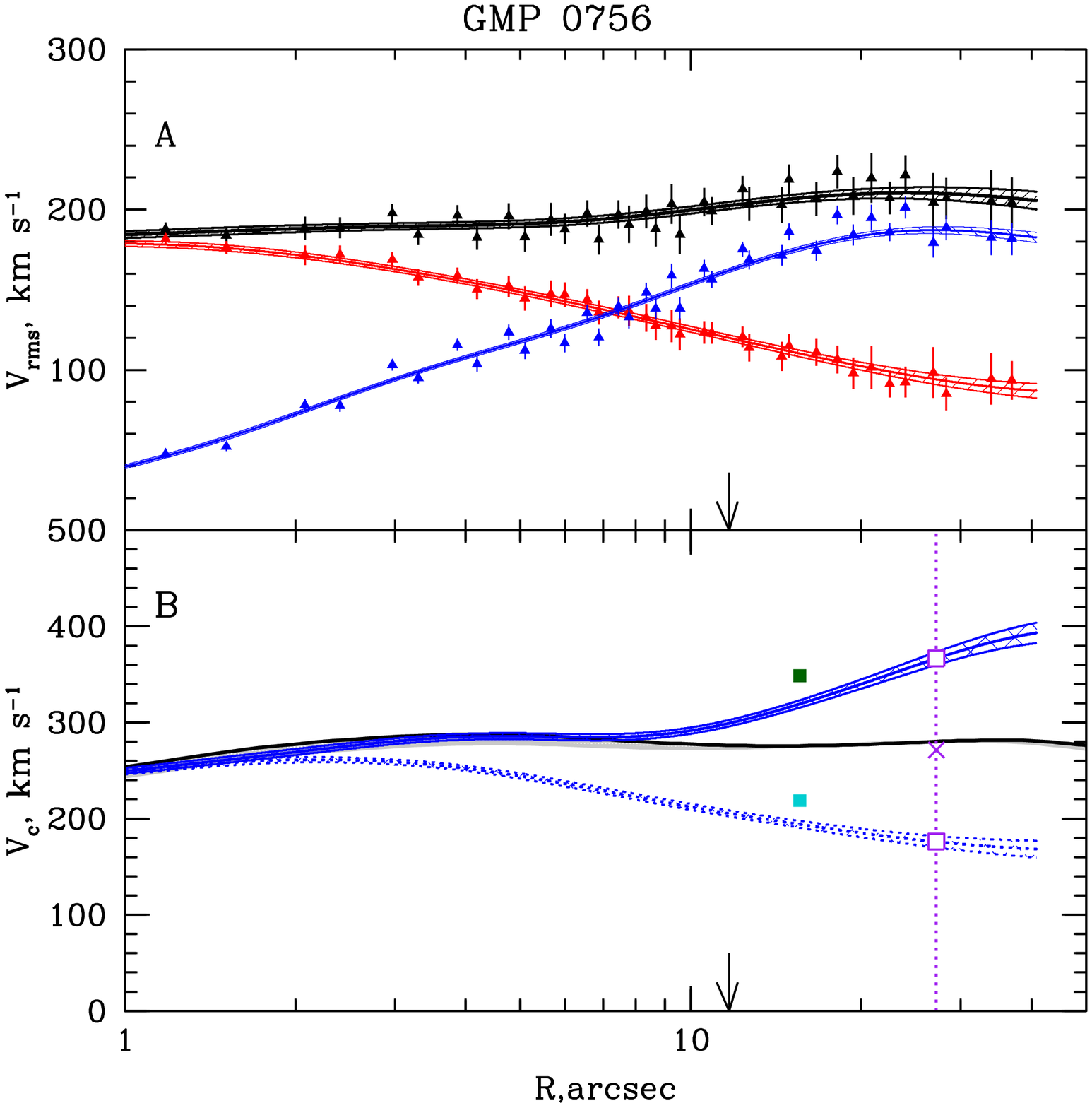}{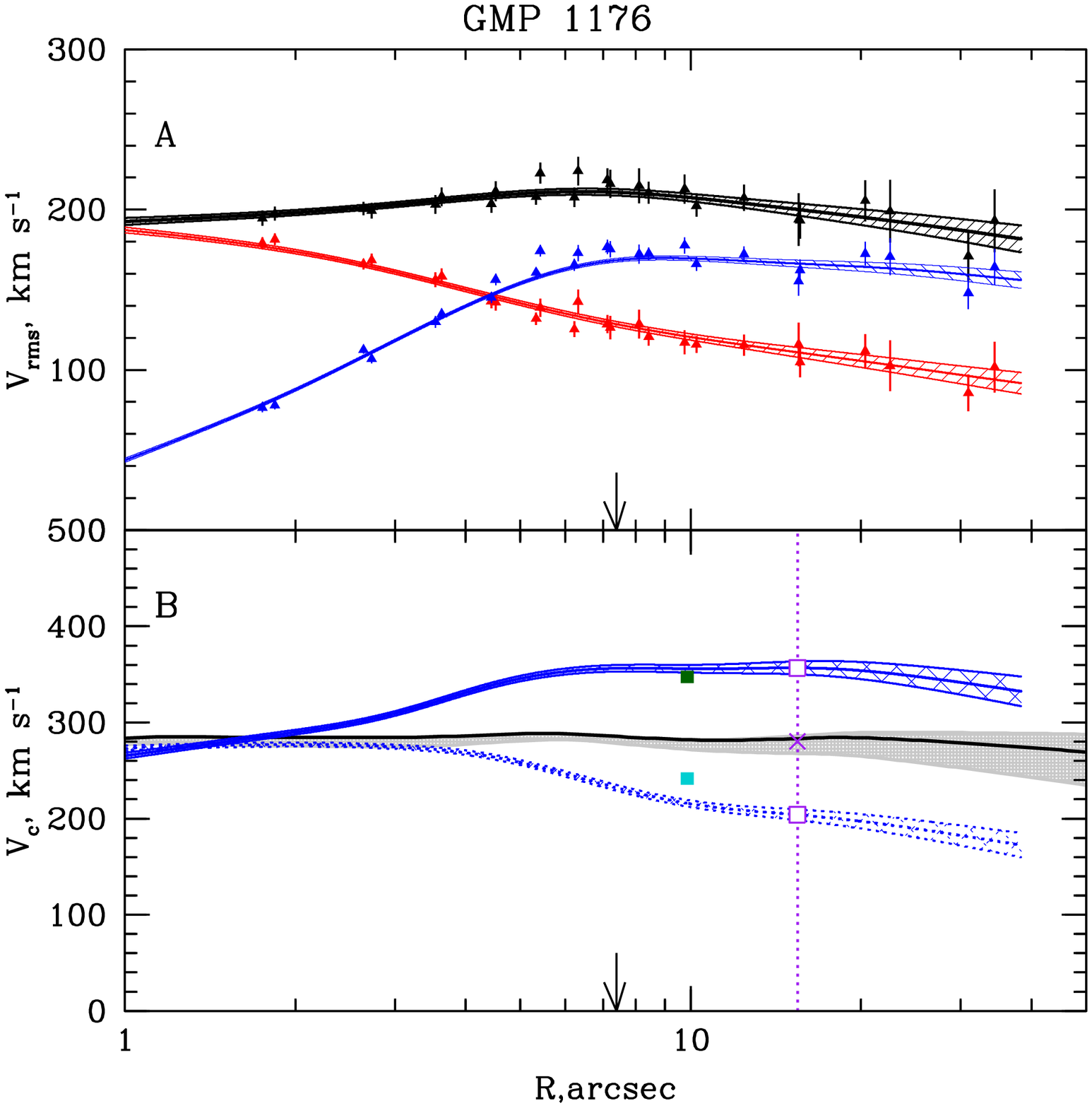}
\plottwos{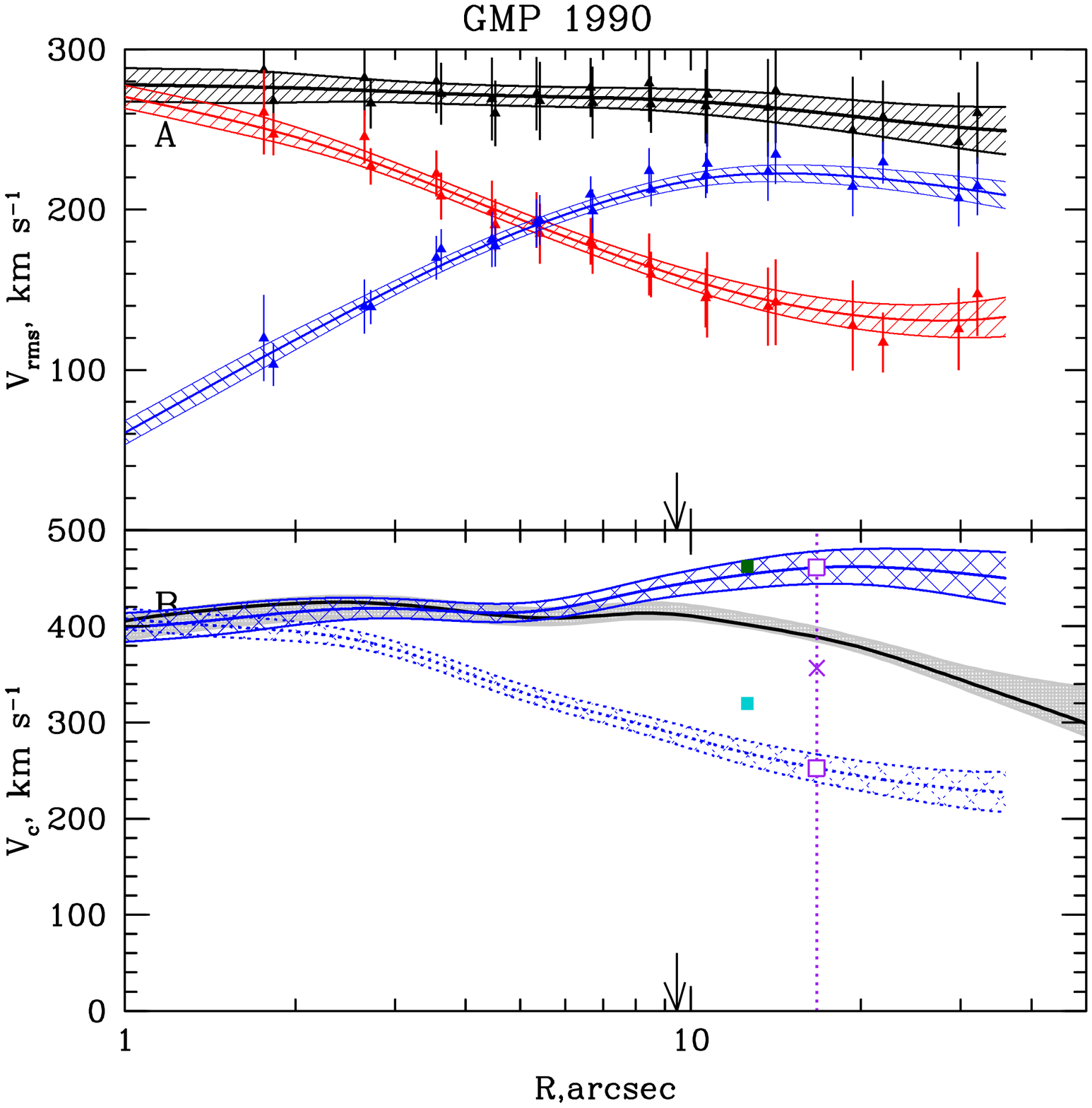}{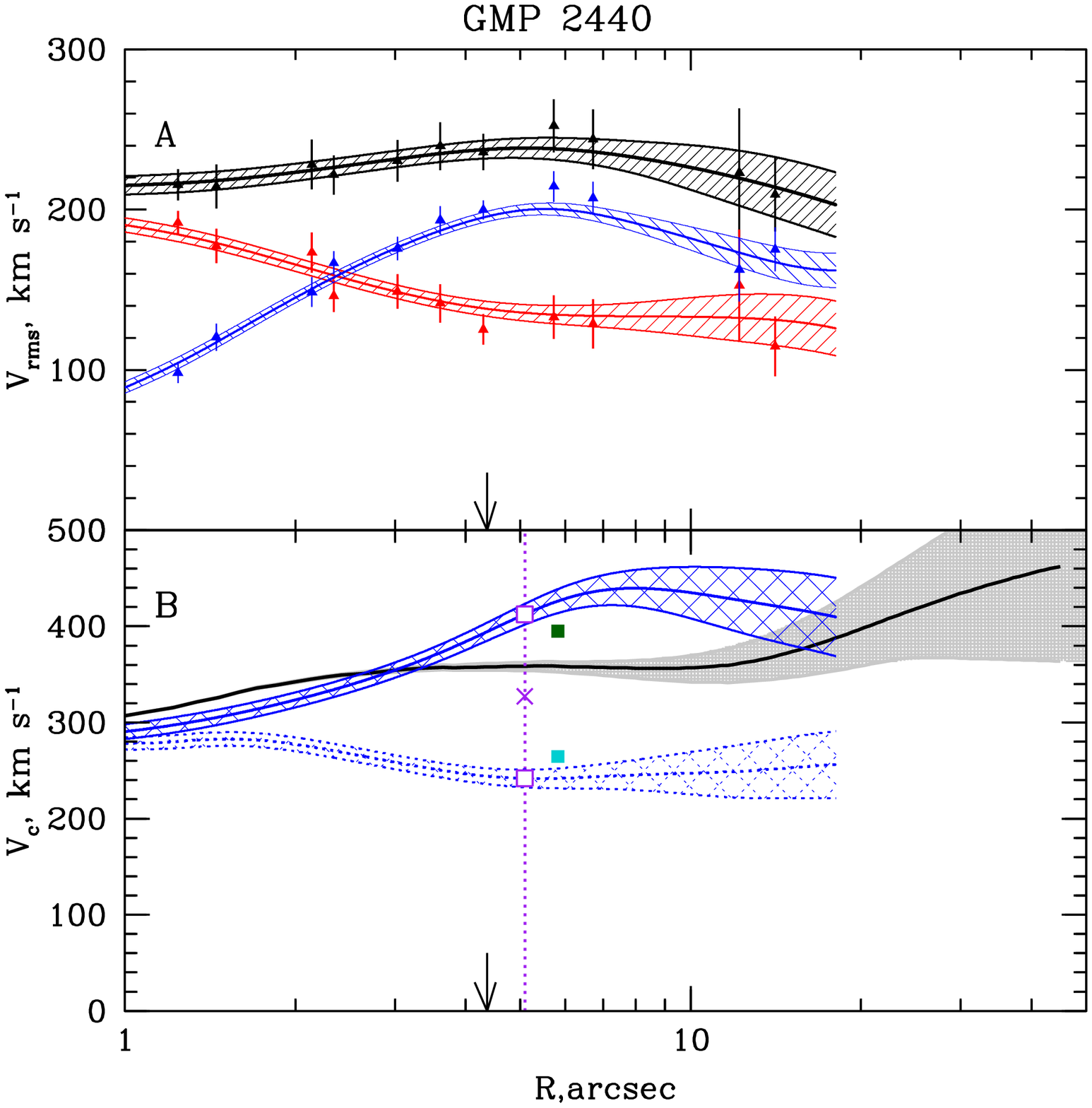}
\caption{Comparison of simple $V_c$-estimates with the circular speed ($V_c^{Schw}$) coming from the dynamical modeling for four galaxies from \citet{Thomas.et.al.2007b} which  have kinematics available out to $\gtrsim 1.5 R_{\rm eff}$ and $\sigma_p(R_{\rm eff})<V_{rot}(R_{\rm eff})$. Panel A shows $ V_{RMS}(R)=\sqrt{\sigma_p^2(R)+V_{rot}^2(R)}$ (in black), $ \sigma_p(R)$ (in red) and $ V_{rot}(R)$ (in blue) profiles measured along the major axis as well as interpolated curves used to calculate the logarithmic derivatives. 
Panel B presents the circular speed resulting from the Schwarzschild modeling (black thick curve) with error bars (grey shaded region), the isotropic circular speed calculated from the RMS velocity $ V^{\rm iso}_c=V_{RMS}\sqrt{1+\alpha+\gamma}$ (in blue, solid lines) and  the isotropic $V_c$ which does not account for rotation $ V^{\rm iso}_c=\sigma_{p}\sqrt{1+\alpha+\gamma}$ (in blue, dotted lines). The simple local $V_c$-estimates at $R_2$ (where $ \alpha = d \ln I/ d \ln R = 2$) are shown as open purple squares, the simple global estimates at $ \frac{4}{3} R_{\rm eff}$ - as filled squares. The arrow and the dotted purple line show the effective radius $R_{\rm eff}$ and $R_2$ respectively. The purple cross marks the average between simple local estimates.} 
\label{fig:art_rot}
\end{figure*}

\begin{figure*}
\plottwo{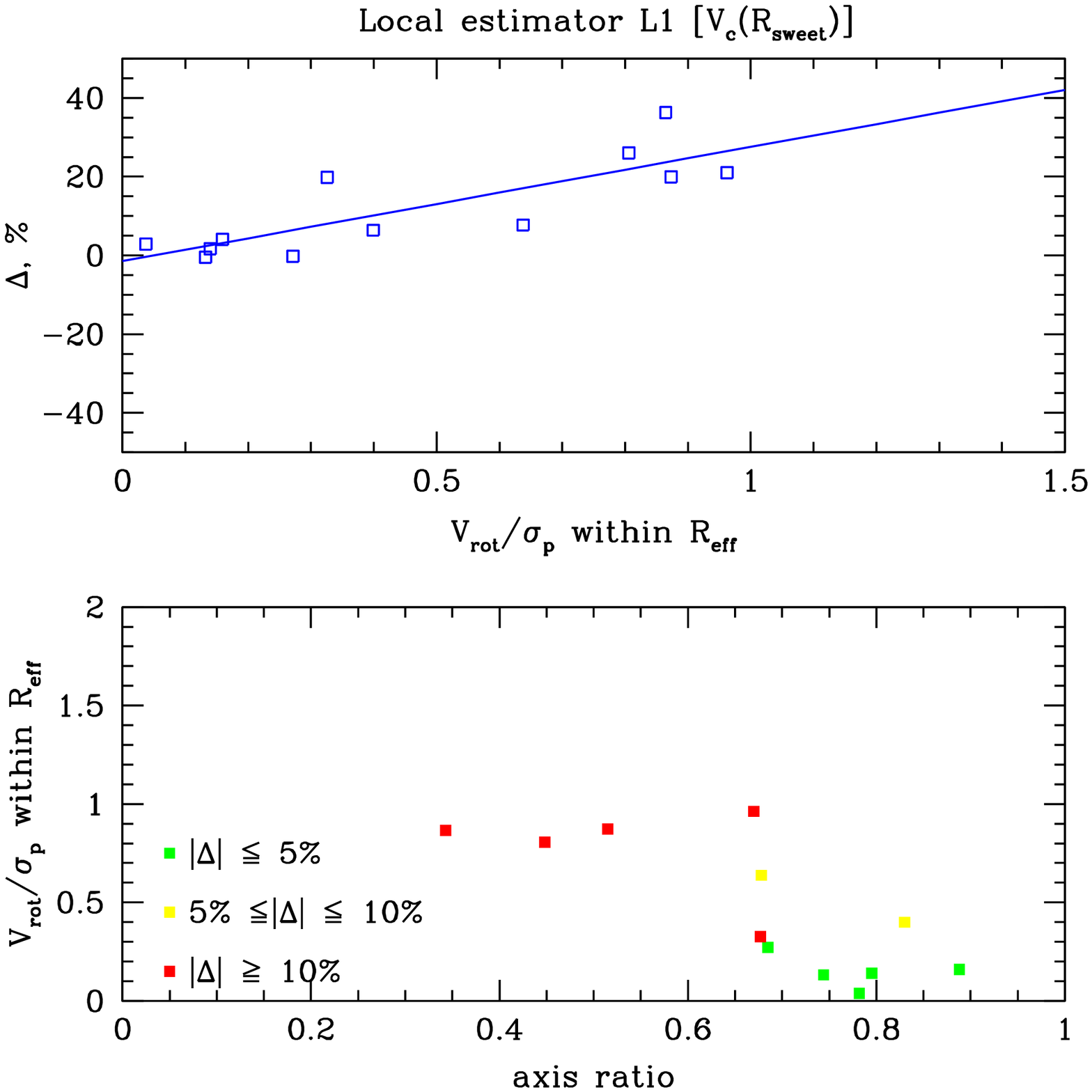}{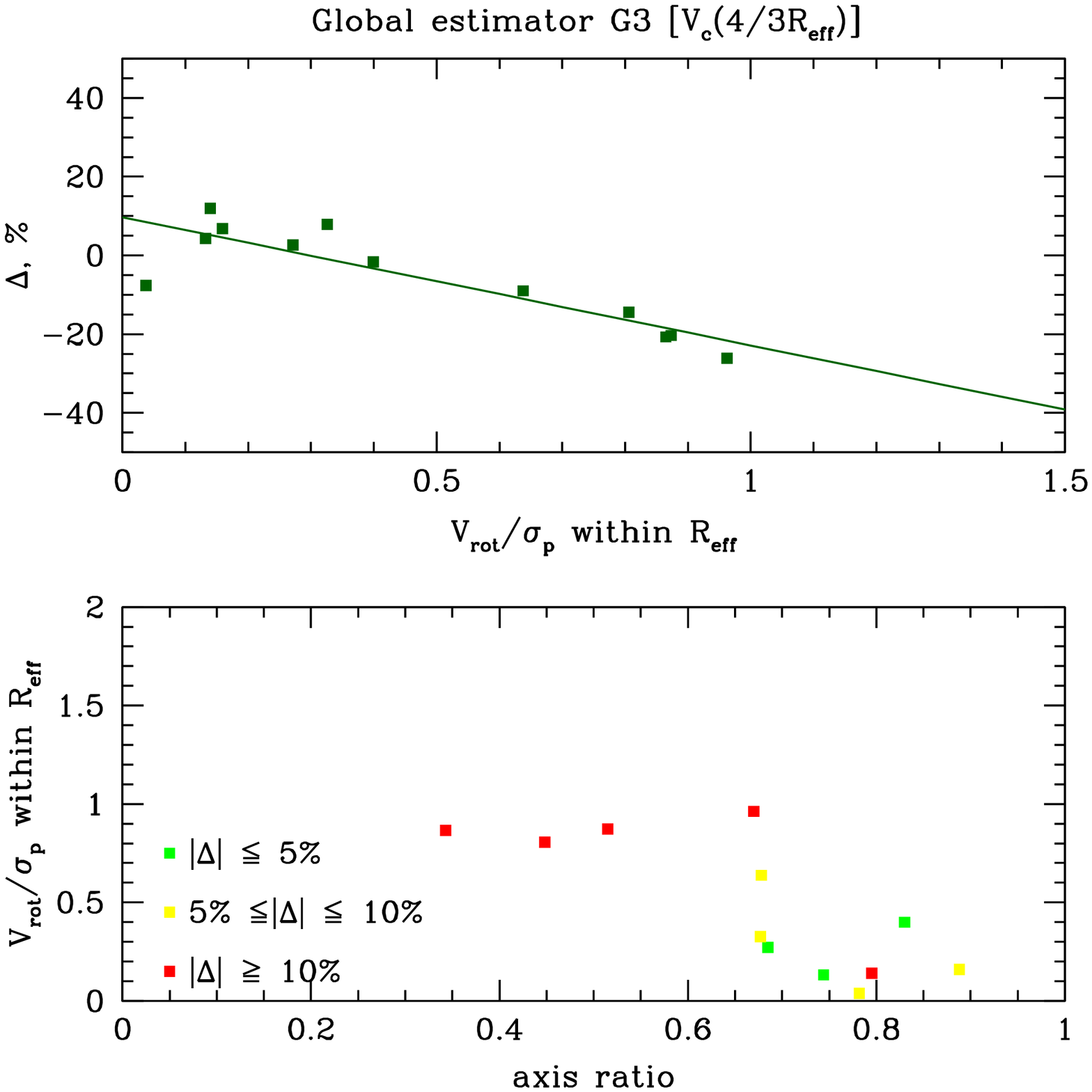}
\caption{Influence of rotation on the simple $V_c$-estimates derived for the sample of real early-type galaxies (11 from \citealt{Thomas.et.al.2007b} plus M87). Deviation $\Delta$ as a function of $ V_{rot}/\sigma_p$, measured within $R_{\rm eff}$, is shown on the upper row. The lower row presents the color-coded $\Delta$  as a function of $ V_{rot}/\sigma_p$ and axis ratio.} 
\label{fig:art_rot2}
\end{figure*}

For galaxies with a significant amount of rotation (i.e., $V_{rot}(R_{\rm eff})>\sigma_p(R_{\rm eff})$) simple methods provide notably biased $V_c$-estimates. Figure~\ref{fig:art_rot} illustrates the comparison of simple estimates with the Schwarzschild dynamical modeling of four additional early-type galaxies from the Thomas et al. sample, which have kinematics along the major axis available out to $\gtrsim 1.5 R_{\rm eff}$ and $\sigma_p(R_{\rm eff})<V_{rot}(R_{\rm eff})$. When the RMS velocity ($ V_{RMS}= \sqrt{\sigma_p^2+V_{rot}^2}$) is used for deriving the circular speed of a galaxy, then the final $V_c$-estimate overestimates the `true' circular speed (coming from dynamical modeling) by $15-30\%$. When one does not account for rotation and uses only projected velocity dispersion for estimating $V_c$, then the final result is biased low by $15-30 \%$. The truth seems to lie somewhere in between. The $V_c$-estimate inferred from $V_{RMS}$ is expected to be almost unbiased only after averaging over galaxies with random 
inclinations \citep{Lyskova.et.al.2014}. The investigated galaxies with a significant amount of rotation are mainly oblate galaxies seen edge-on \citep[see][]{Thomas.et.al.2007b}. And it is not surprising that for such objects with measured kinematics along the major axis only the estimator in a form of $ V_c^2(R_{char})=k\sqrt{\sigma_p^2+ \xi V_{rot}^2}$ with $ \xi \approx 0.5$  would give more sensible $V_c$-estimate. As a crude approximation, we can assume that along the minor axis the rotation would vanish and the velocity dispersion measurements remain almost the same. Then averaging $ V_{RMS}$ over major and minor axes would result in $ V_{RMS} \approx \sqrt{\sigma_p^2+0.5V_{rot}^2}$ where $\sigma_p$ and $V_{rot}$ are measured along the major axis.

Rotation seems to be the main factor which drives the bias of simple circular speed estimates. For this sample of real early-type galaxies (11 from \citealt{Thomas.et.al.2007b} plus M87) the deviation $(V_c-V_c^{Schw})/V_c^{Schw}$ correlates with the luminosity-weighted ratio $V_{rot}/\sigma_{p}$ measured within the effective radius surprisingly well (Figure~\ref{fig:art_rot2}).  


\section{Mass proxy}
\label{sec:proxy}

\begin{figure*}
\plottwos{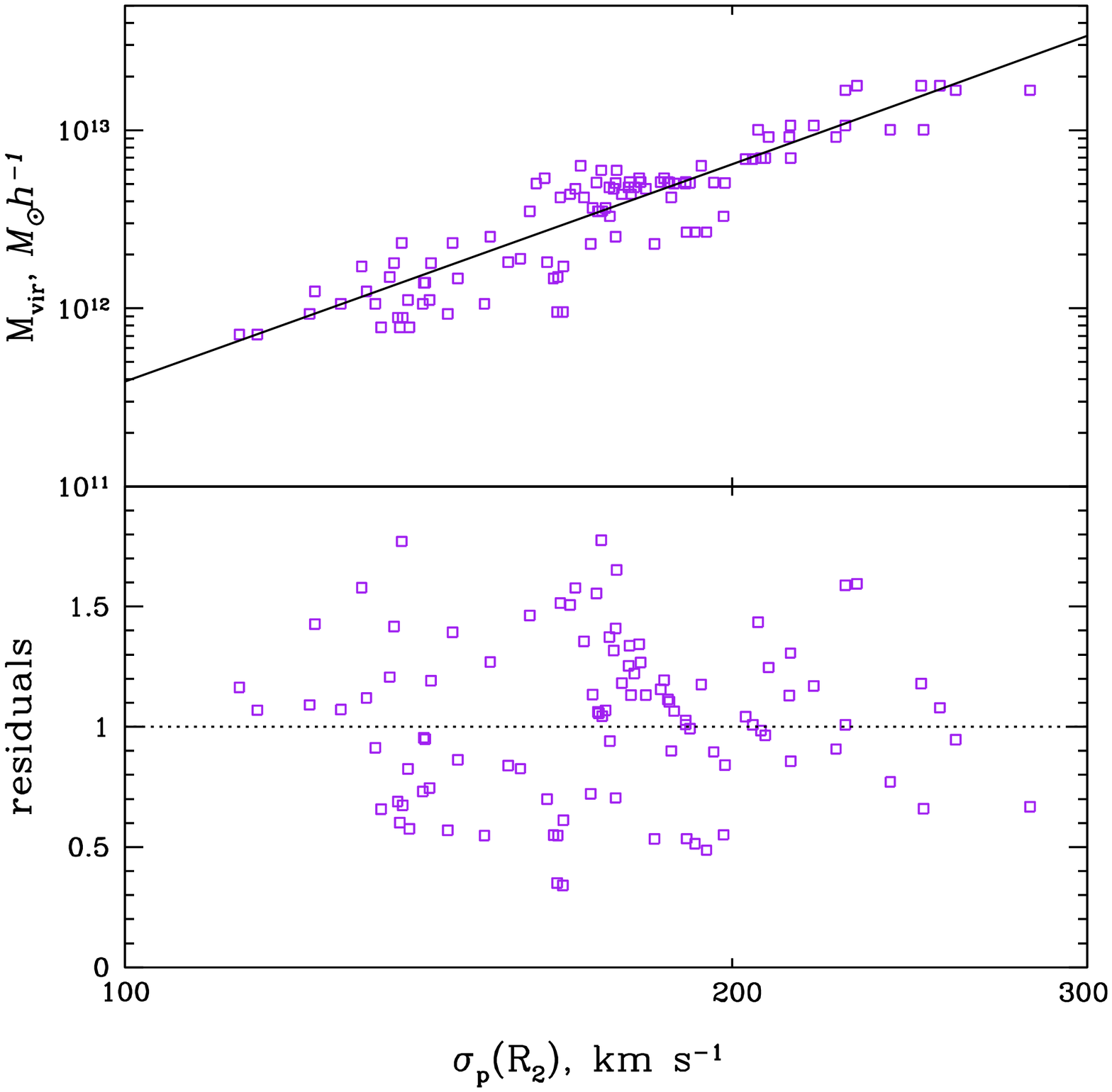}{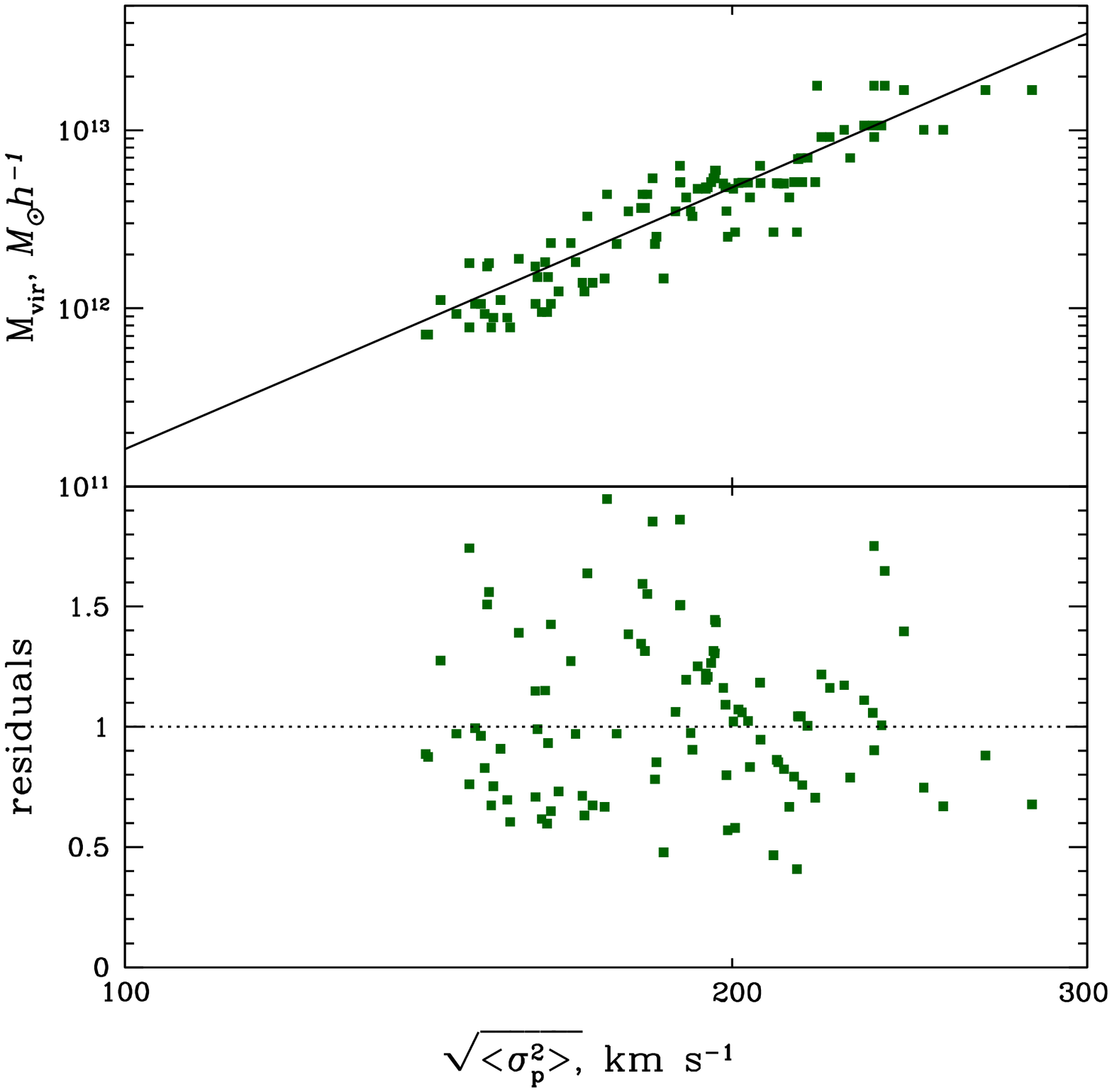}
\caption{The virial mass of simulated galaxies as a function of the projected velocity dispersion at $R_2$ (left) and the luminosity-weighted RMS velocity dispersion (right).} 
\label{fig:proxy}
\end{figure*}

We also test on simulated galaxies whether the simple circular speed estimates could be used as a proxy for the virial galaxy mass.  The simple estimators rely on $\sigma_p(R_2)$ and $\langle \sigma_p^2 \rangle$. Let us plot these against the virial halo mass for each individual galaxy to see how well these quantities correlate with each other.

Figure~\ref{fig:proxy} shows the virial galaxy mass as a function of the local value of the projected velocity dispersion $R_2$ (left side) and the luminosity-weighted average velocity dispersion  $ \langle \sigma_p^2 \rangle^{1/2}$ (right side). Colored squares depict the velocity dispersion and the virial mass for individual simulated galaxies (upper panels) and the straight line is the least-square log-linear fit to these data points. Deviations from the linear fit are shown in the lower panels. The virial mass $M_{vir}$ (in $M_{\odot}h^{-1}$) can be approximated by 
$\disp M_{vir}  \approx 6.0\cdot 10^{12} \left( \frac{\sigma_p(R_2)}{200\, \rm km\, s^{-1}} \right)^{4}$ with RMS $ \approx 38 \%$ or $\disp M_{vir} \approx  4.5 \cdot 10^{12} \left( \frac{\sqrt{\langle\sigma_p^2 \rangle}}{200\rm \, km\, s^{-1}} \right)^{5}$ with RMS $ \approx 37 \%$. The dispersion measures $ \sigma_p(R_2)$ and  $ \langle \sigma_p^2 \rangle $ predict halo mass with almost the same scatter. $ \langle \sigma_p^2 \rangle $  is expected to serve as a good proxy for the virial halo 
mass for our sample of simulated galaxies as the majority of them have almost logarithmic gravitational potential which is well approximated by  $ \Phi(r) \approx 3 \langle \sigma_p^2 \rangle \ln r + const$ at any radius according to the virial theorem. Curiously, the relation between supermassive black hole mass $ M_{BH}$ and host-galaxy bulge velocity dispersion $ M_{BH} \propto \sigma^{4.24} $ \citep{Gultekin.et.al.2009} has approximately the same scatter ($ \approx 31 \%$) and the same exponent as the $ M_{vir}-\sigma_p(R_2)$ and $ M_{vir}-\langle \sigma_p^2 \rangle $ relations.

We also test whether $ \sigma_p(R_{1/2})$ and the luminosity-weighted projected velocity dispersion $ \langle \sigma_p^2 \rangle_e$  measured within an aperture of radius $R_{1/2}$ correlate with the virial mass. We find that the RMS scatter for $ \sigma_p(R_{1/2})$ is $\approx 60 \%$ and for $ \langle \sigma_p^2 \rangle_e$ the scatter is $\approx 50 \%$, i.e., noticeably larger than for $ \sigma_p(R_2)$ and $ \langle \sigma_p^2 \rangle$. 
We have not experimented to determine whether $ \sigma_p(R_2)$ and $ \langle \sigma_p^2 \rangle$ are the best predictors of the halo mass. Therefore, we do not exclude that velocity dispersion measurement at/within other radii could work better. Note that in contrast to the estimators discussed in Section \ref{sec:mass_app}, the halo mass estimators are empirical results only obtained for the sample of massive isolated simulated galaxies, with no dynamical justification.

\section{Discussion and Conclusions}
\label{sec:conclusion}

\begin{table*}
\centering
\caption{\label{tab:summary} Comparing the performance of different estimators for different samples. Best-perfoming estimators are shown in bold face. }
\begin{tabular}{lccccc}
\hline
\multicolumn{1}{l}{}
 &  \multicolumn{2}{c}{\bf Local estimator}
 & \multicolumn{3}{c}{\bf Global estimator} \\
\hline
\multicolumn{1}{l}{Applicable to}
 &  \multicolumn{2}{c}{slowly rotating elliptical galaxies}
 &  \multicolumn{3}{c}{slowly rotating stellar systems} \\
\multicolumn{1}{l}{}
 &  \multicolumn{2}{c}{}
 &  \multicolumn{3}{c}{with nearly flat $\sigma_p(R)$} \\
\hline
\multicolumn{1}{l}{}
 &  \multicolumn{2}{c}{log-slope of $I(R)$ near $R_{char}$,}
 &  \multicolumn{3}{c}{deprojection of $I(R)$} \\
\multicolumn{1}{l}{Data}
 &  \multicolumn{2}{c}{$\sigma_p(R_{char})$,}
 &  \multicolumn{3}{c}{or determination of $R_{1/2}$,} \\
\multicolumn{1}{l}{}
 &  \multicolumn{2}{c}{log-slope of $\sigma_p(R)$ near $R_{char}$ and}
 &  \multicolumn{3}{c}{$\sigma_p(R)$ over entire galaxy} \\
\multicolumn{1}{l}{}
 &  \multicolumn{2}{c}{$d^2\ln[I\sigma_p^2]/d(\ln R)^2$ (for L1)}
 &  \multicolumn{3}{c}{} \\
\hline
\multicolumn{1}{l}{$R_{char}$}
 &  \multicolumn{2}{c}{$R_{\rm sweet}$ or $R_2$}
 &  \multicolumn{3}{c}{$r_3$ or $r_{1/2}$ or $\frac{4}{3}R_{1/2}$} \\
\multicolumn{1}{l}{$\left( V_c^2(R_{char})=k\sigma_p^2\right)$}
 &  \multicolumn{2}{c}{}
 &  \multicolumn{3}{c}{}\\
\hline
\multicolumn{1}{l}{$\sigma_p$}
 &  \multicolumn{2}{c}{$\sigma_p(R_{char})$} 
 &  \multicolumn{3}{c}{$\sqrt{\langle \sigma_p^2 \rangle}$}\\
\hline
\multicolumn{1}{l}{$k$}
 &  \multicolumn{2}{c}{$ 1+\alpha(R_{char})+\gamma(R_{char})$}
 &  \multicolumn{3}{c}{$3$ } \\
\multicolumn{1}{l}{}
 &  \multicolumn{2}{c}{or $ 1+\alpha(R_{2})$}
 &  \multicolumn{3}{c}{} \\
\hline
\rule{0cm}{0.4cm}
 Estimator               & L1 $\disp [V_c(R_{\rm sweet})]$ & L2 $\disp  [V_c(R_{2})]$ & G1 $\disp [V_c(r_{3})]$ & G2 $\disp [V_c(r_{1/2})]$ & G3 $\disp [V_c(\frac{4}{3}R_{1/2})]$\\
\cline{2-6}
\multicolumn{1}{l}{}
&  \multicolumn{5}{c}{spherical analytical models}\\
\cline{2-6}
Average deviation $\disp \bar{\Delta}, \%$&$\bf 1.78 \pm 0.02$             & $2.50 \pm 0.03$    &  $-4.00 \pm 0.03$     &   $-2.75 \pm 0.04$     &  $-2.49 \pm 0.04$ \\ 
$RMS, \% $        & {\bf 3.66}                      & 4.48               &  5.56               &    6.59              &  7.01          \\
\cline{2-6}
\multicolumn{1}{l}{}
&  \multicolumn{5}{c}{simulated galaxies}\\
\cline{2-6}
Average deviation $\disp \bar{\Delta}, \%$& $\bf 0.0 \pm 0.8$             & $-0.1 \pm 0.8$    &  $0.7 \pm 1.3$     &   $0.1 \pm 1.6$     &  $3.4 \pm 1.7$ \\ 
$RMS, \%$         & {\bf 5.4}                     & 5.6               &  6.9               &    9.6              &  7.9           \\
\cline{2-6}
\multicolumn{1}{l}{}
&  \multicolumn{5}{c}{real elliptical galaxies}\\[0.1cm]
\cline{2-6}
Average deviation $\disp \bar{\Delta}, \%$&  $\bf 5.2 \pm 2.3$        & $3.3 \pm 2.7$    & $\bf 2.6 \pm 2.1$                   &   $1.3 \pm 3.0$                  & $2.0 \pm 2.6$  \\ 
$RMS, \%$         & {\bf 6.6}                      & 7.5               &  {\bf  6.0}                &  8.5                   &  7.4           \\
\hline
\end{tabular}
\end{table*}

We have compared the performance of two simple and fast methods  to
evaluate masses of elliptical galaxies at a special radius where
the mass estimate is largely insensitive to the  anisotropy in velocity dispersion. 
Such methods could be useful for mass determination of large samples of galaxies with
poor/noisy data when detailed investigation is not practical. A reliable mass estimate at a single radius could also be used as an additional constraint for the Schwarzschild dynamical modeling thus reducing the range of gravitational potentials to be explored.

One approach uses local properties of the galaxy - logarithmic slopes (and sometimes curvature) of the surface brightness and velocity dispersion profiles and recovers the mass at a radius where the surface brightness declines as $R^{-2}$ (local estimator; \citealt{Churazov.et.al.2010}). Another approach uses the total luminosity-weighted velocity dispersion $\langle \sigma^2_p \rangle $ and evaluates the mass at the radius where the 3D luminosity density $ j(r) \propto r^{-3}$ which can be related to 3D and projected half-light radii as $ r_{3} \approx r_{1/2} \approx \frac{4}{3} R_{1/2}$ for a wide range of stellar distributions (global estimator; \citealt{Wolf.et.al.2010}).    
We test the accuracy and robustness of these simple mass estimators on analytical models, 
on a sample of cosmological zoom-simulations of individual elliptical galaxies
and on real elliptical galaxies that have already been analyzed by means of a 
Schwarzschild approach.

We have found that:

(i) for the analytical models both methods recover the true circular speed with high accuracy. For a grid of explored models the average deviation of the simple local $V_c$-estimate from the true circular speed is $\bar{\Delta} \simeq 2\%$ with RMS scatter $\simeq 4\%$. The global method gives $\bar{\Delta} \simeq -4\%$ with $\simeq 6\%$ scatter. Although the exact values of the average deviation and the RMS scatter depend on the sampling of the parameter space,  the local estimator seems to be less sensitive to the assumptions under which it has been derived than the global one.

(ii) We also examined massive ($\sigma_p(R_{1/2})>150$ km~s$^{-1}$) simulated galaxies, excluding oblate objects seen almost along the rotation axis  ($\sim 15\%$ of the total). For these the local formulae L1, L2 recover an (almost) unbiased estimate of the circular speed with RMS scatter $\approx 5-6 \%$. The Wolf et al. relation also gives an almost unbiased measurement of $V_c$ at the radius where the 3D luminosity density declines as $ r^{-3}$, with RMS $ \approx 7 \%$. At $\frac{4}{3} R_{\rm eff}$ (where the effective radius $R_{\rm eff}$ is defined from the S\'{e}rsic fit to the surface brightness profile) the average global circular speed estimate is biased high by $3.4 \%$ and the RMS scatter is around $8 \%$. For real elliptical galaxies $R_{\rm eff}$ is subject to additional uncertainty (especially when the S\'ersic index is large) as its determination depends on the radial range used for the analysis and applied methodology  and this is another advantage of the local 
estimator. 

(iii) For a sample of eight real slowly-rotating elliptical galaxies with $\sigma_p(R_{\rm eff})<V_{rot}(R_{\rm eff})$ analyzed with the Schwarzschild approach both methods show a remarkable agreement with the best-fit circular speed coming from the dynamical modeling. When averaged over the sample of eight galaxies our simple estimator overestimates the best-fit dynamical circular speed by $5 \%$. This bias is mostly driven by a single galaxy (IC 3947) with the smallest $R_{\rm eff}$ in the sample.  When this galaxy is excluded from the sample, the $V_c(R_{sweet})$-estimator gives a circular speed estimate that is high on average by $3.2 \%$ relative to the $V_c^{Schw}$ with RMS scatter $\approx 3.1 \%$.  The RMS scatter between our simple estimates for the investigated sample of eight galaxies is $\approx 6.6 \%$ which is comparable to measurement uncertainties. The Wolf et al. estimator for the same sample gives a mean overestimate $\approx 2 \%$ with slightly larger RMS scatter of 
$\approx 7.4 \%$.

(iv) A galaxy appearing round on the sky could also be an intrinsically flattened system (e.g., an oblate galaxy viewed along the polar axis), in which case the simple estimates  (and the Schwarzschild models) are expected to become less accurate. For samples of massive ellipticals, the contamination with flattened, face-on galaxies is expected to be small \citep[e.g.,][]{Emsellem.et.al.2011}.

Table~\ref{tab:summary} provides the average deviation and the RMS scatter for different simple estimators resulting from the tests on spherical analytical models, simulated and real elliptical galaxies. Note that the local estimators at $R_{\rm sweet}$ and $R_2$  
show almost the same perfomance, suggesting that instead of searching for a radius $R_{\rm sweet}$ by solving eq.~(\ref{eq:main}) one can use the radius $R_2$, where the surface brightness declines as $R^{-2}$, as the characteristic radius.

The projected velocity dispersion value at the radius $R_2$ where the surface brightness declines as $R^{-2}$ seems to be a good proxy for the virial galaxy mass. $M_{vir}$ (in $M_{\odot} h^{-1}$) can be approximated by $\disp M_{vir}  \approx 6\cdot 10^{12} \left( \frac{\sigma_p(R_2)}{200\, \rm km\, s^{-1}} \right)^{4}$ with RMS scatter $\approx 40 \%$. The scatter is comparable to the scatter observed when $\sqrt{\langle \sigma_p^2 \rangle}$ is used as a proxy for the virial halo mass.

While we were writing this paper, we came across a paper of \cite{Agnello.et.al.2014}, who suggest, in particular, an `extension' of the local approach for power-law total density profiles ($ \rho_{tot} \propto r^{-a}$) in the form of $ V_c^2(R_M) = K\sigma_p^2(R_{\sigma})$, where  $R_{\sigma}$ is a radius where dependence on the exponent $a$ is minimal, $R_{M}$ is chosen in a way to minimize dependence of $R_{\sigma}$ on the anisotropy profile and $K$ is a dimensionless constant.  For example, in case of the S\'ersic surface brightness with $n=4$ and the Osipkov-Merritt anisotropy profile the circular speed of a galaxy can be estimated as $V_c(3.4R_{\rm eff}) \simeq 1.67 \sigma_p(1.15 R_{\rm eff})$. 
To apply the simple mass estimator proposed by \cite{Agnello.et.al.2014} to our grid of analytical models and the sample of simulated galaxies, we derived triplets ($R_{\sigma}, R_{M}, K$) for the corresponding  S\'ersic surface brightness fits and the Osipkov-Merritt 
anisotropy profiles $ \beta(r) = r^2/(r^2+r_a^2)$ with $r_a = 1R_{\rm eff},3R_{\rm eff}$ and $10R_{\rm eff}$. For the sample of analytical models the average deviation is found to be $\simeq -7.4 \% $ and the RMS scatter is $\simeq 3.4 \%$. The negative bias is also found for the investigated simulated galaxies: $\bar{\Delta} = -10\%$, RMS $\simeq 5\%$. For real early-type galaxies the circular speed profile derived from the Schwarzschild modeling at large radii ($R \sim 3R_{\rm eff}$) becomes quite uncertain making a comparison of the estimated $ V_c(R_M)$ at $R_m \gtrsim 3 R_{\rm eff}$ with the `true' one not informative. So from the tests on spherical analytical models and simulated galaxies we conclude that the simple estimator $ V_c^2(R_M) = K\sigma_p^2(R_{\sigma})$ seems to have no obvious advantages over the discussed global and local approaches.

\section{Acknowledgments} 

We are grateful to the referee for very useful comments and suggestions.
NL is grateful to the International Max Planck Research School on Astrophysics (IMPRS) for financial support. This research was supported in part by NASA grant NNX11AF29G. 

\vspace{1cm}

\label{lastpage}
\end{document}